\documentclass[twocolumn,english,aps,prb,twocolum,superscriptaddress,bibnotes,amsmath,amssymb,floatfix]{revtex4-1}
\usepackage[colorlinks=true,citecolor=blue,linkcolor=magenta]{hyperref}

\usepackage[markup=nocolor, authormarkupposition=left]{changes} 
\usepackage{soul}
\usepackage[utf8]{inputenc}
\usepackage[english]{babel}
\usepackage{amsmath,amsfonts,amssymb} 
\usepackage[T1]{fontenc}
\usepackage{url}
\usepackage{makecell}

\usepackage{amsmath}
\usepackage{amsfonts}
\usepackage{amssymb}

\usepackage{epstopdf}
\usepackage{graphicx}
\graphicspath{{./Figures/}}

\usepackage{changes}

\begin{document}
\title{An alkali-referenced vector spectrum analyzer for visible-light integrated photonics}

\author{Baoqi Shi}
\thanks{These authors contributed equally to this work.}
\affiliation{International Quantum Academy, Shenzhen 518048, China}
\affiliation{Department of Optics and Optical Engineering, University of Science and Technology of China, Hefei 230026, China}

\author{Ming-Yang Zheng}
\thanks{These authors contributed equally to this work.}
\affiliation{Jinan Institute of Quantum Technology and CAS Center for Excellence in Quantum Information and Quantum Physics, University of Science and Technology of China, Jinan 250101, China}
\affiliation{Hefei National Laboratory, University of Science and Technology of China, Hefei 230088, China}

\author{Yunkai Zhao}
\affiliation{International Quantum Academy, Shenzhen 518048, China}
\affiliation{Shenzhen Institute for Quantum Science and Engineering, Southern University of Science and Technology, Shenzhen 518055, China}

\author{Yi-Han Luo}
\affiliation{International Quantum Academy, Shenzhen 518048, China}

\author{Jinbao Long}
\affiliation{International Quantum Academy, Shenzhen 518048, China}

\author{Wei Sun}
\affiliation{International Quantum Academy, Shenzhen 518048, China}

\author{Wenbo Ma}
\affiliation{Jinan Institute of Quantum Technology and CAS Center for Excellence in Quantum Information and Quantum Physics, University of Science and Technology of China, Jinan 250101, China}

\author{Xiu-Ping Xie}
\affiliation{Jinan Institute of Quantum Technology and CAS Center for Excellence in Quantum Information and Quantum Physics, University of Science and Technology of China, Jinan 250101, China}
\affiliation{Hefei National Laboratory, University of Science and Technology of China, Hefei 230088, China}

\author{Lan Gao}
\affiliation{International Quantum Academy, Shenzhen 518048, China}
\affiliation{Shenzhen Institute for Quantum Science and Engineering, Southern University of Science and Technology, Shenzhen 518055, China}

\author{Chen Shen}
\affiliation{International Quantum Academy, Shenzhen 518048, China}
\affiliation{Qaleido Photonics, Shenzhen 518048, China}

\author{Anting Wang}
\affiliation{Department of Optics and Optical Engineering, University of Science and Technology of China, Hefei 230026, China}

\author{Wei Liang}
\affiliation{Suzhou Institute of Nano-tech and Nano-bionics, Chinese Academy of Sciences, Suzhou 215123, China}

\author{Qiang Zhang}
\affiliation{Jinan Institute of Quantum Technology and CAS Center for Excellence in Quantum Information and Quantum Physics, University of Science and Technology of China, Jinan 250101, China}
\affiliation{Hefei National Laboratory, University of Science and Technology of China, Hefei 230088, China}
\affiliation{Hefei National Research Center for Physical Sciences at the Microscale and School of Physical Sciences, University of Science and Technology of China, Hefei 230026, China}
\affiliation{CAS Center for Excellence in Quantum Information and Quantum Physics, University of Science and Technology of China, Hefei 230026, China}

\author{Junqiu Liu}
\email[]{liujq@iqasz.cn}
\affiliation{International Quantum Academy, Shenzhen 518048, China}
\affiliation{Hefei National Laboratory, University of Science and Technology of China, Hefei 230088, China}

\maketitle
\noindent\textbf{Integrated photonics has reformed our information society by offering on-chip optical signal synthesis, processing and detection with reduced size, weight and power consumption. 
As such, it has been successfully established in the near-infrared (NIR) telecommunication bands. 
With the soaring demand in miniaturized systems for biosensing, quantum information and transportable atomic clocks, extensive endeavors have been stacked on translating integrated photonics into the visible spectrum, i.e. visible-light integrated photonics.  
A variety of innovative visible-light integrated devices has been demonstrated, such as lasers, frequency combs, and atom traps, highlighting the capacity and prospect to create chip-based optical atomic clocks that can make timing and frequency metrology ubiquitous. 
A pillar to the development of visible-light integrated photonics is characterization techniques featuring high frequency resolution and wide spectral coverage, which however remain elusive. 
Here, we demonstrate a vector spectrum analyzer (VSA) for visible-light integrated photonics, offering spectral bandwidth from 766 to 795 nm and frequency resolution of 415 kHz. 
The VSA is rooted on a widely chirping, high-power, narrow-linewidth, mode-hop-free laser around 780 nm, which is frequency-doubled from the near-infrared via an efficient, broadband CPLN waveguide. 
The VSA is further referenced to hyperfine structures of rubidium and potassium atoms, enabling 8.1 MHz frequency accuracy.
We apply our VSA to showcase the characterization of loss, dispersion and phase response of passive integrated devices, as well as densely spaced spectra of mode-locked lasers. 
Combining operation in the NIR and visible spectra, our VSA allows characterization bandwidth exceeding an octave, and can be an invaluable diagnostic tool for spectroscopy, nonlinear optical processing, imaging and quantum interfaces to atomic devices.
}
%%%%%%%%%%%%%%%%%%%%%%%%%%%%%%%%%%%%%%%%%%%%%%%%%%%%%%%%%%%%%%%%%%%%

\noindent \textbf{Introduction}. 
Integrated photonics \cite{Thomson:16}, which utilizes established semiconductor manufacturing technology for construction and mass production of chip-scale optical systems, has made explosive growth in the last decades. 
Enabling dense integration of lasers, modulators and detectors on monolithic substrates for optical information processing, integrated photonics has revolutionized telecommunications, sensing and computing.  
As such, it has been extensively developed and optimized in the NIR wavelength, e.g. around 1550 nm where today's telecommunications and datacenters operate \cite{Agrell:16}.   

\begin{figure*}[t!]
\centering
\includegraphics[width=\linewidth]{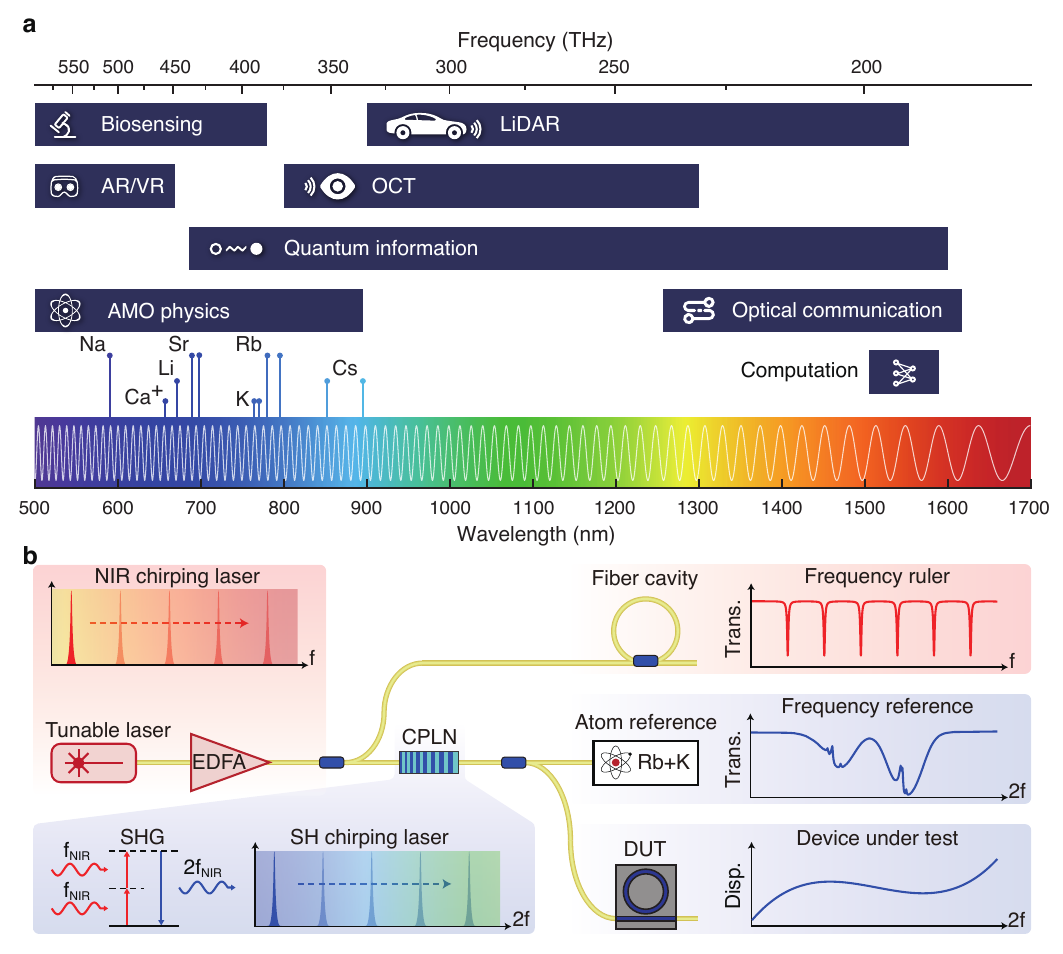}
\caption{
\textbf{Principle and applications of vector spectrum analyzers}.
\textbf{a}. 
Applications of integrated photonics in the NIR and visible spectra.
\textbf{b}.
Principle and schematic of our alkali-referenced vector spectrum analyzer in the visible spectrum.
The NIR chirping laser is a telecommunication-C-band, widely tunable, CW laser.
After power amplification by an EDFA, a CPLN waveguide is used to frequency-double the NIR laser and generate the SH chirping laser.
The relative frequency of both chirping lasers is calibrated by an FSR-calibrated fiber cavity that serves as a time-domain ``frequency ruler''.
The absolute frequency of both chirping lasers is referenced to Rb and K hyperfine structures.
The frequency-calibrated SH chirping laser is used to perform vector spectrum analysis of a device under test (DUT) in the visible spectrum.
}
\label{Fig:1}
\end{figure*}

Currently, there is surging interest in translating integrated photonics into the visible spectrum \cite{Blumenthal:20, He:20, Tran:22, Poon:24}. 
Figure \ref{Fig:1}a presents diverse applications operated in the visible spectrum, such as biosensing \cite{Helle:20, Ettabib:24}, augmented and virtual reality (AR/VR) \cite{ChangC:20, Xiong:21}, AMO (atomic, molecular and optical) physics \cite{Ludlow:15, Pezze:18}, LiDAR (light detection and ranging) \cite{KimI:21, Li:22}, OCT (optical coherence tomography) \cite{Yun:17, Bouma:22}, and quantum information \cite{Gisin:07, Kimble:08}. 
A notable application is chip-based atomic and optical clocks \cite{Kitching:18, Knappe:04, Papp:14, Newman:19}. 
Despite that transportable clocks with precision reaching $10^{-18}$ have been demonstrated \cite{Koller:17, Masao:20}, chip-based atomic and optical clocks bear further reduced size, weight and power consumption, and can allow frequency metrology ubiquitously deployed on mobile platforms and in space. 
Endeavors have created integrated components dedicated to clock systems, including low-noise lasers \cite{Chauhan:21, Franken:21, Siddharth:22, Corato-Zanarella:23, Li:23, Clementi:23, Ling:23}, frequency combs \cite{Li:17, Pfeiffer:17, Yu:19, Chen:20}, surface-electrode ion traps \cite{Mehta:20, Niffenegger:20}, and magneto-optic traps \cite{Isichenko:23}. %LiM:22,
Recently, integrated low-noise lasers have been deployed to interrogate strontium (Sr) ion clocks \cite{Loh:24, Chauhan:24}. 

%%%%%%%%%%%%%%%%%%%%%%%%%%%%%%%%%%%%%%%%%%%%%%%%%%%%%%%%%%%%%%%%%%%%
%%%%%%%%%%%%%%%%%%%%%%%%%%%%%%%%%%%%%%%%%%%%%%%%%%%%%%%%%%%%%%%%%%%%

Parallel to the progress in new architecture, fabrication, and application, characterization techniques of integrated photonics in the visible spectrum are equally pivotal, which however are inadequately developed.
Wavemeter-calibrated tunable lasers and Mach–Zehnder interferometers \cite{Desiatov:19, Chauhan:21, Choy:12} have been employed to quantify optical loss or resonance linewidth of integrated microresonators. 
However these methods require individual measurement for each resonance, and lack accuracy to determine resonance frequency. 
Cutback measurement can evaluate optical loss based on long waveguides with varying lengths \cite{Morin:21, Franken:21, Bradley:12, Hegeman:20}, however it suffers from mediocre precision and is invalid for cavity structures. 
None of these methods have sufficient measurement bandwidth to map the dispersion profiles, mainly due to the absence of widely tunable, narrow-linewidth, mode-hop-free lasers in the visible spectrum. 

%%%%%%%%%%%%%%%%%%%%%%%%%
Here, we demonstrate a wideband, high-resolution vector spectrum analyzer (VSA) for visible-light integrated photonics. 
Our VSA operates in the 766 to 795 nm wavelength -- the spectral region dedicated to rubidium optical clocks. 
The principle of our VSA is illustrated in Fig. \ref{Fig:1}b.
While commercial continuous-wave (CW) lasers in the visible spectrum have mode hopping and limited range during chirping, here we use a widely tunable, mode-hop-free, external-cavity diode laser (Santec TSL) in telecommunication band, which is frequency-doubled to the visible spectrum.
The chirp range of this NIR laser is from 1532 to 1590 nm (7.1 THz), and the chirp rate is set to 50 nm/s. 
The laser is amplified by an erbium-doped fiber amplifier (EDFA), and is split into two branches.
One branch is directed to a fiber cavity with a quasi-equidistant grid of fine resonances.
The fiber cavity's free spectral range (FSR) is calibrated using sideband modulation driven by a microwave source, as illustrated in Ref. \cite{Luo:24}. 
Therefore, when the NIR laser chirps, the fiber cavity's transmission spectrum containing an FSR-calibrated resonance grid serves as a time-domain ``frequency ruler'' (see Supplementary Materials Note 1 for details). 
The relative laser frequency $f_\mathrm{NIR,r}$, e.g. the frequency difference from the initial laser frequency or a reference frequency (discussed later), is calibrated on-the-fly by referring to this ``frequency ruler''.

The other branch is coupled into a chirped periodically poled lithium niobate (CPLN) waveguide.
Harnessing broadband quasi-phase-matching (QPM) \cite{Umeki:09, Ycas:18, Wu:22, Wu:24} in the CPLN waveguide, the NIR laser's frequency is doubled into the visible spectrum via second-harmonic generation (SHG). 
As such, we create a CW, widely tunable, mode-hop-free laser in the visible spectrum, whose chirp bandwidth is also doubled to 14.3 THz (766 to 795 nm).
Therefore, the second-harmonic (SH) laser's relative frequency is calculated as $f_\mathrm{SH,r}=2f_\mathrm{NIR,r}$. 
It is apparent that relative frequency calibration of $f_\mathrm{NIR,r}$ in the NIR is equivalent to that of $f_\mathrm{SH,r}$ in the visible spectrum.

The SH laser is then split into two branches.
One branch is directed to an atomic reference, which comprises two vapor cells containing rubidium (Rb) and potassium (K), respectively. 
Employing saturated absorption spectroscopy \cite{Preston:96}, the hyperfine structures of these alkali atoms provide absolute frequency references.
Consequently, with relative frequency calibration provided by the fiber cavity and the absolute frequency calibration by the alkali atoms, the instantaneous frequency of the SH laser ($f_\mathrm{SH}$), as well as the NIR laser ($f_\mathrm{NIR}$), is precisely and accurately calibrated during laser chirping. 
The other branch of the SH laser is directed to a device under test (DUT), whose loss, dispersion, and phase response in visible spectrum are characterized.

%%%%%%%%%%%%%%%%%%%%%%%%%%%%%%%%%%%%%%%%%%%%%%%%%%%%%%%%%%%%%%%%%%%
\begin{figure*}[t!]
\centering
\includegraphics[width=\linewidth]{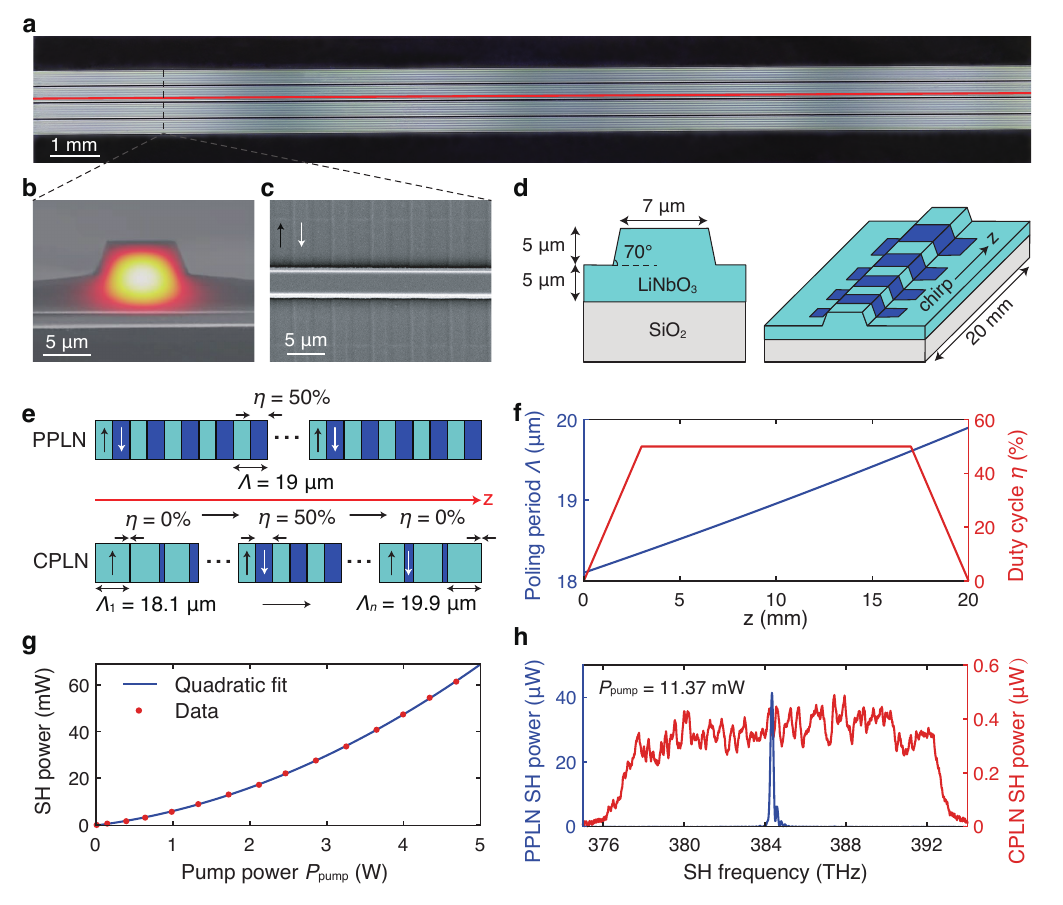}
\caption{
\textbf{Chirped periodically poled lithium niobate waveguide and second-harmonic laser generation}.
\textbf{a}. 
Optical microscope image of the CPLN waveguide chip.
The waveguide is outlined with red color.
\textbf{b}.
SEM image showing the CPLN waveguide cross-section with overlaid TM$_{00}$ optical mode.
\textbf{c}.
SEM image of the CPLN waveguide.
Different polarization directions are marked by black and white arrows.
The poling period is seen. 
\textbf{d}.
Illustration of waveguide geometry.
\textbf{e}.
Illustration of poling period $\Lambda$ and duty cycle $\eta$ of the PPLN and CPLN waveguides.
\textbf{f}.
Design of poling period $\Lambda$ and duty cycle $\eta$ of the CPLN waveguide over 20 mm length.
\textbf{g}.
Second-harmonic power generated by the CPLN waveguide, with different pump power $P_\mathrm{pump}$ at 1560 nm. 
\textbf{h}.
Bandwidth comparison of the SH lasers generated by the PPLN and CPLN waveguides, respectively.
The NIR pump power is $P_\mathrm{pump}=11.37$ mW. 
}
\label{Fig:2}
\end{figure*}
%%%%%%%%%%%%%%%%%%%%%%%%%%%%%%%%%%%%%%%%%%%%%%%%%%%%%%%%%%%%%%%%%%%%

\vspace{0.3cm}
%%%%%%%%%%%%%%%%%%%%%%%%%%%%%%%%%%%%%%%%%%%%%%%%%%%%%%%%%%%%%%%%%%%%
\noindent \textbf{Second-harmonic laser}. 
Critical to converting the NIR laser to the SH laser is broadband, efficient SHG. 
Periodically poled lithium niobate (PPLN) waveguides based on QPM have been widely employed due to their high efficiency, ultrafast response, and compact size \cite{Yamada:93}. 
In QPM gratings, the phase mismatch between the interacting optical waves, i.e. the pump and the SH, can be compensated by the cyclically inverted nonlinear coefficient, as shown in Fig. \ref{Fig:2}e. 
Exploiting the large nonlinear optical coefficient $d_{33}$ of lithium niobate (LiNbO$_3$), PPLN devices empower efficient SHG within its transparency window from 400 to 5000 nm.

However, a fixed poling period $\Lambda$ of PPLN limits the SH bandwidth up to hundreds of gigahertz, unfavorable in our case.
To extend the SH bandwidth, we design and fabricate a CPLN waveguide with chirped $\Lambda$ on a z-cut, MgO-doped, thin-film LiNbO$_3$ wafer \cite{Zhu:21}. 
The fabrication process is elaborated in Supplementary Materials Note 2.
Figure \ref{Fig:2}a shows the optical microscope image of the CPLN waveguide.
Figure \ref{Fig:2}(b,c,d) shows the cross-sectional and longitudinal structures of the CPLN waveguide.
The LiNbO$_3$ ridge waveguide has 7 $\mu$m top width, 5 $\mu$m etching depth, 10 $\mu$m total thickness, 70$^{\circ}$ sidewall angle, and $20$ mm length.
The waveguide is designed such that the NIR pump laser is fiber-coupled to the waveguide's fundamental-magnetic (TM$_{00}$) mode, as shown in Fig. \ref{Fig:2}b. 
Figure \ref{Fig:2}(e,f) shows that, along our 20-mm-long CPLN waveguide, $\Lambda$ is varied from 18.1 to 19.9 $\mu$m. 
Meanwhile, the duty cycle $\eta$, i.e. the ratio of the inversely poled waveguide length in one poling cycle, is increased from 0\% to 50\% in the first 3 mm length, and decreased from 50\% to 0\% in the last 3 mm.
The duty cycle $\eta=50\%$ remains in the middle of the waveguide.
The duty-cycle engineering reduces the fluctuation of SH conversion efficiency over a wide bandwidth (see Supplementary Materials Note 2).

We use the CPLN waveguide for SHG without an EDFA to characterize the generated SH bandwidth and measure the SH power. 
The output SH laser from the CPLN waveguide is coupled into a single-mode fiber for 780 nm wavelength, where the remaining NIR laser is dissipated.
The fiber is connected to a photodetector that probes the laser power. 
The measured SH power over 375 to 394 THz is shown in Fig. \ref{Fig:2}h red curve.
The 3-dB spectral bandwidth is 15.1 THz (377.4 THz to 392.5 THz).
In comparison, a PPLN waveguide with identical geometry but $\Lambda=19$ $\mu$m and $\eta=50\%$, as shown in Fig. \ref{Fig:2}e, is also fabricated and characterized. 
Figure \ref{Fig:2}h blue curve shows SH bandwidth generated by the PPLN waveguide.  
Limited to the constant $\Lambda$, the 3-dB bandwidth is only 166 GHz.

In addition to the SH bandwidth, our SH laser's tuning range is also limited by the EDFA's bandwidth.
We use a wideband EDFA to amplify the NIR laser from 1532 to 1590 nm, and to generate the SH laser from 766 to 795 nm (14.3 THz, from 377.1 THz to 391.4 THz). 
To evaluate the SH laser's practical bandwidth for spectrum analysis, an air-gap Fabry–Pérot (F-P) cavity's transmission spectrum is measured (see Supplementary Materials Note 3 for details).
In the full laser bandwidth, the signal-to-noise ratio of the F-P cavity's resonances is over 22.3 dB.

Figure \ref{Fig:2}g shows the measured NIR-pump-power-dependent SH power at 1560 nm with the EDFA.
The SH laser power reaches 68.6 mW with 5 W NIR pump power, sufficient for spectrum analysis.
We also measure the SH power for an hour with 1.4 W pump power, to examine the SH power's stability. 
The measured root mean square (RMS) of the SH power is less than 0.1\%. 

The SH laser's dynamic linewidth during chirping is critical to our VSA's frequency resolution.
We measure this dynamic linewidth using a self-delayed heterodyne setup (see Supplementary Materials Note 4 for details).
Within 100 $\mu$s time scale, the SH laser's dynamic linewidth is averaged as 415 kHz, setting the lower bound of our VSA's frequency resolution.

%%%%%%%%%%%%%%%%%%%%%%%%%%%%%%%%%%%%%%%%%%%%%%%%%%%%%%%%%%%%%%%%%%%%
\begin{figure*}[t!]
\centering
\includegraphics[width=\linewidth]{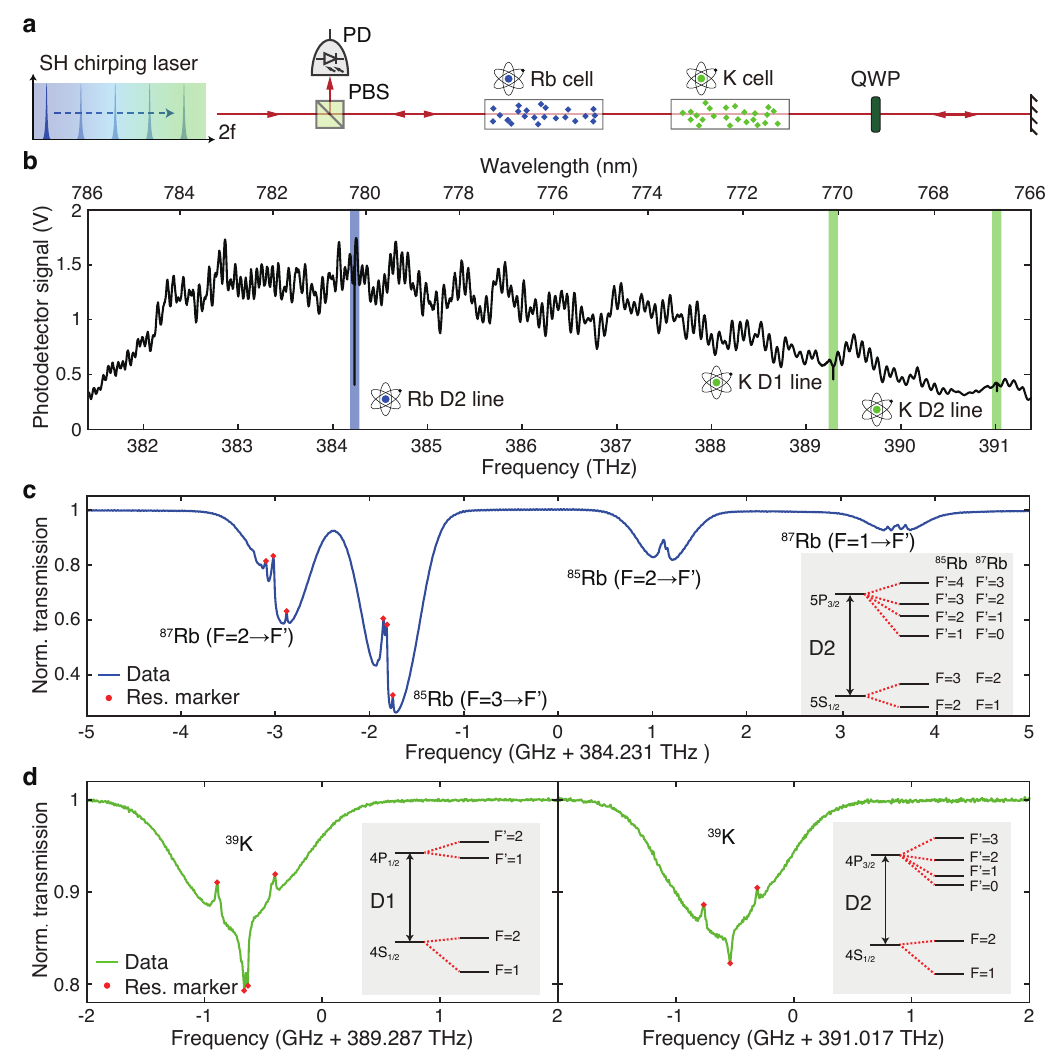}
\caption{
\textbf{Alkali-atom reference and saturated absorption spectroscopy}.
\textbf{a}. 
Schematic of the double-path setup for saturated absorption spectroscopy of Rb and K atoms.
PD, photodetector.
PBS, polarization beam splitter.
QWP, quarter-wave plate.
\textbf{b}.
Saturated absorption spectrum of Rb and K atoms.
The Rb D2 line is blue-shaded. 
The K D1 and D2 lines are green-shaded.
\textbf{c}.
Saturated absorption spectrum of Rb D2 line.
The four Doppler-broadened absorption dips correspond to transition $F=2\rightarrow F'$ of $^{87}$Rb,
$F=3\rightarrow F'$ of $^{85}$Rb,
$F=2\rightarrow F'$ of $^{85}$Rb,
$F=1\rightarrow F'$ of $^{87}$Rb.
The six peaks used for absolute frequency reference of our VSA are marked with red dots.
Inset shows the hyperfine structure of Rb D2 line.
\textbf{d}.
Saturated absorption spectra of K D1 (left) and D2 (right) lines.
The seven peaks used for correction of relative frequency calibration and evaluation of measured frequency accuracy are marked with red dots.
Inset shows the hyperfine structure of K D1 and D2 lines.
}
\label{Fig:3}
\end{figure*}
%%%%%%%%%%%%%%%%%%%%%%%%%%%%%%%%%%%%%%%%%%%%%%%%%%%%%%%%%%%%%%%%%%%%

\vspace{0.3cm}
%%%%%%%%%%%%%%%%%%%%%%%%%%%%%%%%%%%%%%%%%%%%%%%%%%%%%%%%%%%%%%%%%%%%
\noindent \textbf{Alkali-atom frequency reference}. 
Our widely tunable, narrow-linewidth, mode-hop-free SH laser is advantageous to realize saturated absorption spectroscopy \cite{Preston:96} of Rb and K in a single measurement within the laser bandwidth.
The schematic is illustrated in Fig. \ref{Fig:3}a.
Two counter-propagating, spatially overlapped, SH laser beams transmit through two cascaded vapor cells containing Rb and K atoms, respectively.
When the laser frequency resonates with an atomic transition, one beam strongly drives the atomic ensemble, creating minimum absorption for the other beam and thus a narrow transparency window. 
As a result, when the laser frequency chirps across an atomic transition among specific hyperfine levels, a small peak in the Doppler-broadened absorption feature is observed.
The linewidth of the saturated absorption peaks can approach the natural linewidth determined by the atomic energy level's lifetime, e.g. 6 MHz for Rb D2 line \cite{Steck:23}. 
In our measurement, the peak linewidth is around 15 MHz. 
The broadening originates from the power broadening effect \cite{Thomas:77}.

We perform saturated absorption spectroscopy of Rb and K with our SH laser.
The photodetected light signal from the vapor cells is shown in Fig. \ref{Fig:3}b, where Rb D2 line (blue region) and K D1 and D2 lines (green region) are resolved.
Figure \ref{Fig:3}c shows the zoom-in profile of Rb D2 line.  
Six peaks are resolved and marked (red dots), whose information is listed in Methods. 
Referring to the frequency values of these transitions provided in Ref. \cite{Steck:23}, the SH laser's instantaneous frequency $f_\mathrm{SH}$ is determined. 

Figure \ref{Fig:3}d shows the saturated absorption spectra of K D1 and D2 lines. 
The K D1 and D2 lines exhibit four and three peaks, respectively, whose information is listed in Methods.
Using Rb D2 line at $f_\mathrm{Rb}=384.229$ THz for absolute frequency calibration, the three transitions of K D2 line at 391.016 THz are used to examine the accuracy of our relative frequency calibration with the fiber cavity.
A mean deviation of 28.0 MHz is found between our measured values and the values in Ref. \cite{Tiecke:19}. 
This deviation stems from the residual error in calibrating the fiber cavity's FSR and temperature-induced FSR drift.
Therefore, we further correct the SH laser's instantaneous frequency, as 
\begin{equation}
f_\mathrm{SH}'=\frac{f_\mathrm{SH}-f_\mathrm{Rb}}{1+0.000413\%}+f_\mathrm{Rb}
\label{Eq.Correct}
\end{equation}
where $0.000413\%=28.0/(391.016\times10^6-384.229\times10^6)$.
With Eq. \ref{Eq.Correct}, we measure the frequency values of the four peaks of K D1 line at 389.286 THz.
Referred to the values in Ref. \cite{Tiecke:19}, our results exhibit 8.1 MHz mean deviation, corresponding to 1.6 ppm relative frequency error. 
In comparison, a commercial wavelength-meter has a typical frequency accuracy of 200 MHz. 
%%%%%%%%%%%%%%%%%%%%%%%%%%%%%%%%%%%%%%%%%%%%%%%%%%%%%%%%%%%%%%%%%%%%
\begin{figure*}[t!]
\centering
\includegraphics[width=\linewidth]{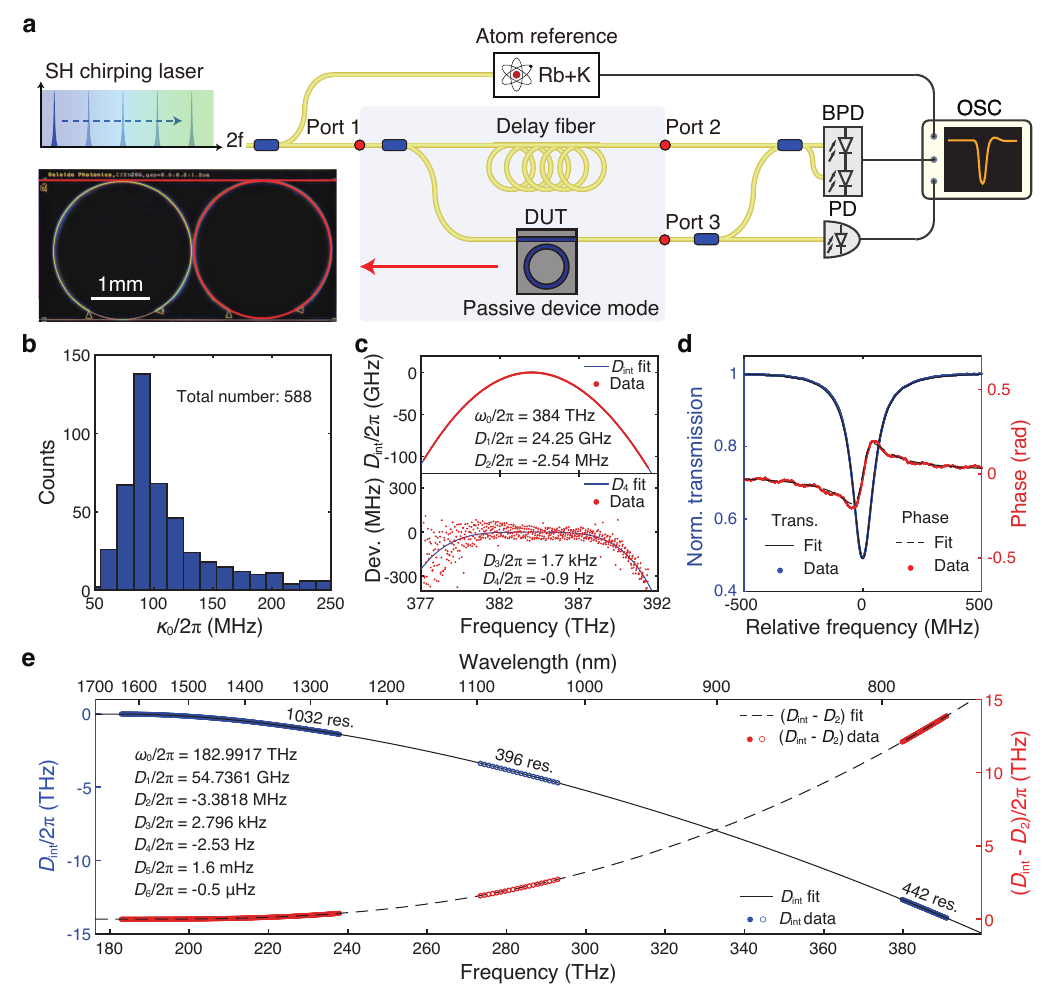}
\caption{
\textbf{Characterization of integrated Si$_3$N$_4$ microresonators}.
\textbf{a}. 
Experimental setup to measure the transmission (loss), dispersion and phase response of Si$_3$N$_4$ microresonators. 
Inset is an optical microscope image showing two Si$_3$N$_4$ microresonators coupled with two bus waveguides, respectively. 
Only one is measured each time, marked with red lines.
BPD, balanced photodetector.
OSC, oscilloscope.
\textbf{b}.
Histogram of measured $\kappa_0/2\pi$ values of the 588 resonances from one microresonator, showing the most probable value of $\kappa_0/2\pi=90$ MHz.
\textbf{c}.
Measured microresonator's integrated dispersion profile $D_\text{int}/2\pi$ and fit up to the fourth order $D_4$.
The deviation is defined as $[D_\text{int}(\mu)-D_2\mu^2/2-D_3\mu^3/6]/2\pi$, showing a clear $D_4$ trend.
\textbf{d}.
Measured transmission and phase profile of one under-coupled resonance.
\textbf{e}.
Measured $D_\text{int}/2\pi$ and $D_3/2\pi$ profiles over octave bandwidth, by combining the SH and NIR chirping lasers.
The $D_\text{int}$ curve is fitted based on 1032 resonances from 1260 to 1640 nm (dots), and 442 resonances from 766 to 795 nm (dots). 
Additional measurement using another VSA operated at 1020.6 to 1096.6 nm wavelength provides 396 resonances (circles), which overlay well with the $D_\text{int}$ curve.
This agreement validates the accuracy and precision of our octave-span $D_\text{int}$ measurement. 
The microresonator dispersion can be fitted up to the sixth order $D_6$.
}
\label{Fig:4}
\end{figure*}
%%%%%%%%%%%%%%%%%%%%%%%%%%%%%%%%%%%%%%%%%%%%%%%%%%%%%%%%%%%%%%%%%%%%
%%%%%%%%%%%%%%%%%%%%%%%%%%%%%%%%%%%%%%%%%%%%%%%%%%%%%%%%%%%%%%%%%%%%
\begin{figure*}[t!]
\centering
\includegraphics[width=\linewidth]{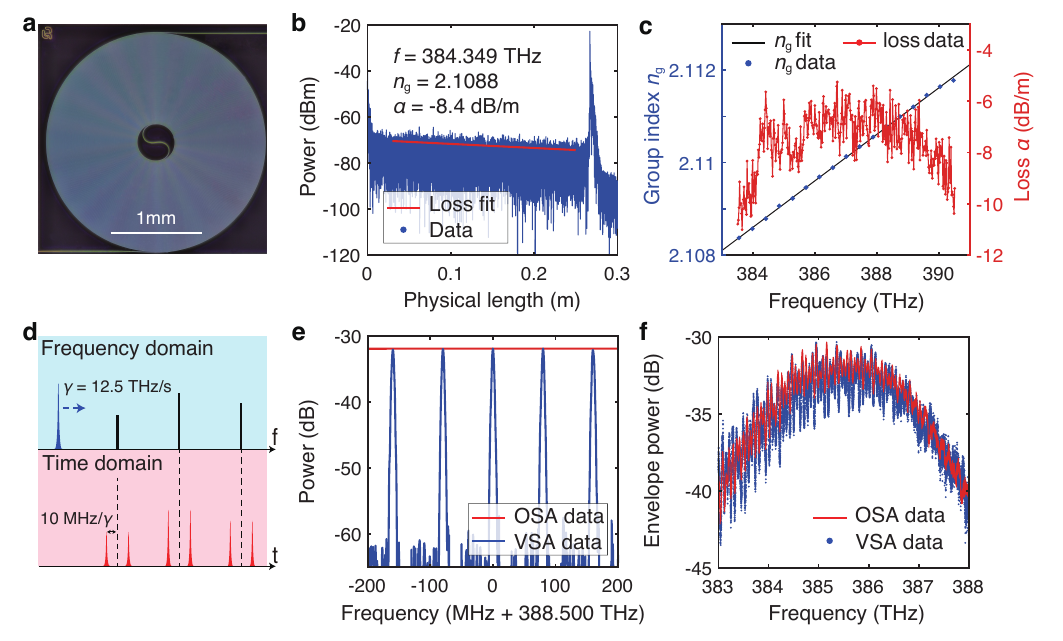}
\caption{
\textbf{Characterization of Si$_3$N$_4$ spiral waveguides and mode-locked laser spectra}.
\textbf{a}. 
Optical microscope image showing a Si$_3$N$_4$ spiral waveguide of 0.26708-meter physical length and $2.5\times2.5$ mm$^2$ footprint. 
\textbf{b}.
Measured OFDR data of the spiral waveguide. 
The major peak at 0.26708 m is due to light reflection at the chip's rear facet, where the waveguide terminates. 
With measured optical length of 0.56322 m, the group index $n_g=2.1088$ at 384.349 THz is calculated.
The loss rate $\alpha=-8.4$ dB/m (physical length) is extrapolated with a linear fit of power decrease over distance (red line).
\textbf{c}.
Measured group index $n_g$ and loss $\alpha$ of the spiral waveguide at different frequencies.
\textbf{d}.
Principle of coherent detection of broadband laser spectra using a chirping CW laser. 
The CW laser beats progressively with different frequency components of the optical spectrum, allowing frequency mapping into the RF domain and continuous information output in the time domain. 
\textbf{e}.
Zoom-in of the mode-locked laser spectrum measured by our VSA (blue) and a commercial OSA (red).
The mode spacing can only be resolved by our VSA. 
\textbf{f}.
Spectral power envelopes measured by our VSA (blue) and the OSA (red).
}
\label{Fig:5}
\end{figure*}
%%%%%%%%%%%%%%%%%%%%%%%%%%%%%%%%%%%%%%%%%%%%%%%%%%%%%%%%%%%%%%%%%%%%

\vspace{0.3cm}
%%%%%%%%%%%%%%%%%%%%%%%%%%%%%%%%%%%%%%%%%%%%%%%%%%%%%%%%%%%%%%%%%%%%
\noindent \textbf{Characterization of passive integrated devices}.
Next, we use our VSA to characterize passive integrated devices in the visible spectrum. 
We first characterize an ultralow-loss, integrated silicon nitride (Si$_3$N$_4$) microresonator. 
The fabrication process is presented in Supplementary Materials Note 5. 
Silicon nitride has 5 eV bandgap, which endows a wide transparency window from ultraviolet to mid-infrared. 
Therefore Si$_3$N$_4$ is a leading platform for integrated photonics in the visible and blue spectra \cite{Hosseini:09, Romero-Garcia:13, Munoz:19, Morin:21, Sanna:24}.
Especially, Si$_3$N$_4$ microresonators are key components to low-noise integrated lasers with visible and blue wavelength \cite{Chauhan:21, Franken:21, Siddharth:22, Corato-Zanarella:23, Li:23, Clementi:23}.
The fabrication and characterization of our Si$_3$N$_4$ microresonators in telecommunication bands are found in Ref. \cite{Ye:23, Luo:24, Sun:24}.
The setup to characterize them in the visible spectrum is depicted in Fig. \ref{Fig:4}a. 
The inset shows an optical microscope image of the Si$_3$N$_4$ chip. 
The SH laser transmits through Port 1 and is edge-coupled into the Si$_3$N$_4$ waveguide's fundamental-electric (TE$_{00}$) mode using a lensed fiber.
Inverse waveguide tapers \cite{Almeida:03} on the chip facets facilitate fiber-waveguide mode match and thus reduce coupling loss (see Supplementary Materials Note 6 for more details). %Ren:11
A polarizer is added before the lensed fiber to achieve 37 dB polarization extinction ratio of the input laser (see Supplementary Materials Note 7 for details).

During laser chirping, the microresonator's transmission is probed by a photodetector through Port 3 and recorded with an oscilloscope.
The microresonator's phase response is calculated by analyzing the time delay between the microresonator and the delay fiber arm \cite{Luo:24}, i.e. time delay between Port 2 and 3.
The microresonator's resonances are fitted using \cite{Aspelmeyer:14, Pfeiffer:18}:
\begin{equation}
s(\Delta\omega)=\frac{\kappa_0-\kappa_\mathrm{ex}+2i\Delta\omega}{\kappa_0+\kappa_\mathrm{ex}+2i\Delta\omega}
\label{Eq.Res}
\end{equation}
where $s$ is the complex amplitude of the microresonator's transmission signal, 
$\Delta\omega/2\pi$ is the laser detuning to the resonance, 
$\kappa_0/2\pi$ is the resonance's intrinsic loss, 
$\kappa_\mathrm{ex}/2\pi$ is the resonance's external coupling strength to the bus waveguide.
The resonance profile is fitted with $\lvert s(\Delta\omega) \rvert^2$ and the phase is extracted by $\arg s(\Delta\omega)$.
A typical resonance with its fit is shown in Fig. \ref{Fig:4}d.
The phase response reveals that the resonance is under-coupled ($\kappa_\mathrm{ex}<\kappa_0$).
The fitted loss values are $\kappa_0/2\pi=89.6$ MHz and $\kappa_\mathrm{ex}/2\pi=16.0$ MHz.
Exploiting the 14.3 THz bandwidth of our VSA, we can identify, measure and fit 588 resonances in one microresonator of 24.25 GHz FSR.
The histogram of $\kappa_0/2\pi$ is shown in Fig. \ref{Fig:4}b, with the most probable value of $\kappa_0/2\pi=90$ MHz, corresponding to an intrinsic quality factor of $Q_0=4.3 \times 10^6$ and a linear loss of 14 dB/m.

The wide bandwidth and high frequency accuracy of our VSA allows mapping the microresonator's integrated dispersion profile, defined as
\begin{equation}
D_\text{int}(\mu)=\omega_\mu-\omega_0-D_1\mu=\sum_{n=2}^{\cdots}\frac{D_n\mu^n}{n!}
\label{Eq.Dint}
\end{equation}
where $\omega_{\mu}/2\pi$ is the $\mu^\text{th}$ resonance frequency relative to the reference resonance frequency $\omega_0/2\pi$,
$D_1/2\pi$ is microresonator FSR, 
$D_2/2\pi$ describes group velocity dispersion (GVD), 
and $D_3$, $D_4$ are higher-order dispersion terms. 
With the 588 resonances from 766 to 795 nm, the dispersion terms up to $D_4$ are measured, as shown in Fig. \ref{Fig:4}c top.
This is validated in Fig. \ref{Fig:4}c bottom, where $D_2$ and $D_3$ are subtracted from $D_\text{int}$. 
The deviations of the data from the fit are likely caused by multi-mode interference \cite{Ji:22}.

We note that, using a portion of the NIR laser, our alkali-referenced VSA can simultaneously characterize the device in the NIR. 
Therefore, we can combine the microresonator dispersion measurements respectively performed in the visible and NIR bands, and obtain an octave-bandwidth dispersion profile. 
In the NIR band, the measurement bandwidth covers from 1260 to 1640 nm \cite{Luo:24}, containing 1032 resonances; 
In the visible spectrum, the measurement bandwidth covers from 766 to 795 nm, containing 442 resonances. 
We infer the relative mode number $\mu$ in the gap between these two bands \cite{YangK:16}.
Figure \ref{Fig:4}e shows the fitted $D_\text{int}$ over an octave bandwidth, whose dispersion terms up to $D_6$ are extrapolated.
In addition, to validate our octave-span $D_\text{int}$ measurement, we build another VSA but operated at 1020.6 to 1096.6 nm wavelength, and characterize the resonances of the same Si$_3$N$_4$ microresonator (see Supplementary Materials Note 8 for details). 
As shown in Fig. \ref{Fig:4}e, the measured 396 resonances overlay with the $D_\text{int}$ curve, proving the accuracy and precision of our octave-span dispersion characterization.
This capability of octave-span dispersion characterization is critical for dispersion engineering \cite{Okawachi:14},  a technique widely used to generate octave-spanning soliton microcomb \cite{Li:17, Pfeiffer:17, Yu:19, Chen:20}, supercontinuum \cite{Zhao:15, Epping:15a, Johnson:15, Porcel:17a}, and spectral translation \cite{Li:16, Lu:19, Billat:17}.

In addition to optical microresonators, our VSA can also function as an optical frequency-domain reflectometry (OFDR) \cite{Soller:05} to characterize long waveguides (see Supplementary Materials Note 9 for details).
Here we measure the loss $\alpha$ and dispersion of a Si$_3$N$_4$ spiral waveguide of 0.26708 m physical length. 
Figure \ref{Fig:5}a shows the optical microscope image of the spiral waveguide of $2.5\times2.5$ mm$^2$ size. 
Figure \ref{Fig:5}b shows the OFDR result measured at 384.349 THz. 
The prominent peak is due to the waveguide's rear facet.
The waveguide's optical length is measured as 0.56322 m.
By dividing the optical length by the physical length, the waveguide's group index $n_\mathrm{g}=2.1088$ is calculated.
Figure \ref{Fig:5}c shows the measured $n_\mathrm{g}$ from 383.540 to 390.459 THz. 
With linear fit, the waveguide dispersion is extrapolated as $\beta_1=7034.1$ fs/mm and $\beta_2=267.2$ fs$^2$/mm at 384.349 THz.

Light traveling in the waveguide experiences attenuation according to the Beer-Lambert law $I(L)=I_0\cdot\text{exp}(\alpha L)$.
By linear fit of the profile envelope of the reflection power (red line in Fig. \ref{Fig:5}b), the average linear loss $\alpha=-8.4$ dB/m (physical length) is extrapolated at 384.349 THz.
Figure \ref{Fig:5}c shows the fitted $\alpha$ value with different frequency.  
The observed fluctuation of $\alpha$ is likely due to multi-mode interference \cite{Ji:22} in the waveguide. 

\vspace{0.3cm}
%%%%%%%%%%%%%%%%%%%%%%%%%%%%%%%%%%%%%%%%%%%%%%%%%%%%%%%%%%%%%%%%%%%%
\noindent \textbf{Characterization of mode-locked laser spectra}.
Besides passive devices, our VSA can also characterize laser spectra.
The setup in Fig. \ref{Fig:4}a is modified by connecting the laser under test to Port 3, and shortcutting Port 1 and 2 (see Supplementary Materials Note 10 for setup).
The emission spectrum of a mode-locked laser (Chameleon Ultra) operated at 775 nm (386.8 THz) is measured simultaneously by our VSA and a commercial optical spectrum analyzer (OSA, Yokogawa AQ6370D).
The laser spectrum consists of tens of thousands of CW modes with $\sim80$ MHz mode spacing. 
For simplicity, we use $f_n$ to denote the $n^\text{th}$ mode's frequency.
When the SH laser chirps at rate $\gamma=12.5$ THz/s, it beats against each mode of the mode-locked laser.
The beat signal is recorded by an oscilloscope and digitally processed by a finite impulse response (FIR) band-pass filter \cite{Shi:24} with 10 MHz passband center frequency and 6 MHz bandwidth (see Supplementary Materials Note 10 for algorithm).
Thus, every time the SH laser chirps across $f_n\pm 10$ MHz, i.e.  $f_\text{SH}'=f_n\pm 10$ MHz, a pair of pulses pass through the FIR bandpass filter, as shown in Fig. \ref{Fig:5}d. 
The pulse pair's amplitude is proportional to the mode power.
Therefore, the mode-locked laser's spectrum is measured by calibrating the SH laser's instantaneous frequency and the pulse power.

Figure \ref{Fig:5}e shows the zoom-in of the laser spectrum measured by the OSA and our VSA.
The OSA has 0.1 nm (50 GHz) resolution, impossible to resolve the fine mode spacing of $\sim80$ MHz.
In contrast, our VSA can provide 3 MHz frequency resolution and thus resolve each mode.
Within the mode-locked laser's spectrum bandwidth from 383 to 388 THz, the power of each mode is extracted and plotted in Fig. \ref{Fig:5}f.  
The spectral power envelopes measured by the OSA and our VSA are nearly identical. 
However, owing to the high frequency resolution, our VSA can resolve the actual contrast of power variation.

\vspace{0.3cm}
%%%%%%%%%%%%%%%%%%%%%%%%%%%%%%%%%%%%%%%%%%%%%%%%%%%%%%%%%%%%%%%%%%%%
\noindent \textbf{Conclusion}.  
In conclusion, we have demonstrated a VSA of 14.3 THz spectral bandwidth in the visible spectrum (766 to 795 nm) and 415 kHz frequency resolution. 
The VSA is based on a widely chirping NIR laser whose frequency is calibrated by a fiber cavity and converted to the visible spectrum via SHG in a CPLN waveguide.
The converted SH laser has 68.6 mW output power, and is further referenced to Rb and K hyperfine structures, endowing 8.1 MHz frequency accuracy to our VSA.
We use our VSA to characterize the loss, dispersion and phase response of passive integrated devices, and densely spaced spectra of mode-locked lasers. 
Particularly, combining simultaneous measurement in the NIR and visible spectra, our VSA allows octave-span dispersion mapping, critical for the development and optimization of wideband integrated nonlinear photonics and quantum interfaces to atomic devices \cite{Spencer:18, Newman:19, Dutt:24}. 

Besides showcasing Si$_3$N$_4$ devices, our VSA is equally essential to the development and translation of visible-light integrated photonics based on LiNbO$_3$ \cite{Desiatov:19}, AlN \cite{LiuX:18a, Lu:18}, TiO$_2$ \cite{Choy:12, Bradley:12, Hegeman:20}, and Al$_2$O$_3$ \cite{West:19, Franken:23}. 
Employing various high-power seed lasers allows bandwidth extension to even shorter wavelengths. 
For example, using the 1064 nm tunable laser, a ytterbium-doped fiber amplifier (YDFA), and a CPLN optimized for 1064 nm, a VSA operated at 532 nm can be created. 
Our work paves a path to ultra-wideband (more than an octave) spectrum analysis using cascaded VSAs operated at different spectral regions, which proves to be an invaluable diagnostic tool for spectroscopy, imaging, sensing and quantum information process.

\vspace{0.3cm}
%%%%%%%%%%%%%%%%%%%%%%%%%%%%%%%%%%%%%%%%%%%%%%%%%%%%%%%%%%%%%%%%%%%%
\noindent \textbf{Methods}. 

\noindent \textbf{Saturated absorption peaks of Rb and K}.
Figure \ref{Fig:3}c shows four Doppler-broadened absorption dips corresponding to the D2 ($5S_{1/2}\rightarrow 5P_{3/2}$) transition of isotopes $^{85}$Rb and $^{87}$Rb.
The four transitions are [$F=2\rightarrow F'$] of $^{87}$Rb,
[$F=3\rightarrow F'$] of $^{85}$Rb,
[$F=2\rightarrow F'$] of $^{85}$Rb,
[$F=1\rightarrow F'$] of $^{87}$Rb.
Red dots mark six saturated absorption peaks. 
For $^{87}$Rb transition [$F=2\rightarrow F'$], the three peaks correspond to [$F=2\rightarrow CO'(1,3)$)], [$F=2\rightarrow CO'(2,3)$], and [$F=2\rightarrow F'=3$], 
Here $CO'(i,j)$ represents the excited-state crossover between $F'=i$ and $F'=j$.
According to Ref. \cite{Steck:23}, these peaks have frequency values of 384.227903407, 384.227981877, and 384.228115203 THz, respectively.
For $^{85}$Rb transition [$F=3\rightarrow F'$], the three peaks correspond to [$F=3\rightarrow CO'(2,4)$], [$F=3\rightarrow CO'(3,4)$], and [$F=3\rightarrow F'=4$].
These peaks have frequency values of 384.229149669, 384.229181369, and 384.229241689 THz, respectively.

Figure \ref{Fig:3}d shows the saturated absorption spectra of K D1 and D2 lines. 
The K vapor cell contains mainly $^{39}$K atoms.
The K D1 transition [$4S_{1/2}\rightarrow 4P_{1/2}$] exhibits four peaks corresponding to [$F=2\rightarrow CO'(1,2)$], [$CO(1,2)\rightarrow CO'(1,2)$], [$CO(1,2)\rightarrow F'=2$], and [$F=1\rightarrow F'=2$], where $CO(1,2)$ denotes the ground-state crossover.
According to Ref. \cite{Tiecke:19}, these four peaks have frequency values of 389.2858787, 389.2861095, 389.2861373, and 389.2863681 THz, respectively.

The K D2 transition [$4S_{1/2}\rightarrow 4P_{3/2}$] exhibits three peaks corresponding to [$F=2\rightarrow F'$], [$CO(1,2)\rightarrow F'$], and [$F=1\rightarrow F'$].
These three peaks have frequency values of 391.0159969, 391.0162278, and 391.0164586 THz, respectively.
%%%%%%%%%%%%%%%%%%%%%%%%%%%%%%%%
\medskip
\begin{footnotesize}

\noindent \textbf{Funding Information}: 
J. Liu acknowledges support from the National Natural Science Foundation of China (Grant No.12261131503), Innovation Program for Quantum Science and Technology (2023ZD0301500), Shenzhen-Hong Kong Cooperation Zone for Technology and Innovation (HZQB-KCZYB2020050), and from the Guangdong Provincial Key Laboratory (2019B121203002).
W. L. acknowledges support from the National Natural Science Foundation of China (62075233), CAS Project for Young Scientists in Basic Research (YSBR-69), and Innovation Program for Quantum Science and Technology (2023ZD0301500).
M. Z. acknowledges support from the Key R\&D Plan of Shandong Province (Grant No. 2021ZDPT01), Natural Science Foundation of Shandong Province (ZR2021LLZ013, ZR2022LLZ009).

\noindent \textbf{Acknowledgments}: 
We thank Zhongkai Wang for the suggestion on the saturated absorption spectrum measurement and Yafei Ding for assistance in taking sample photos. 

\noindent \textbf{Author contribution}: 
B. S., Y. Z., Y.-H. L., J. Long and W. S. built the setup and performed the experiment, with the assistance from W. L. and A. W.. 
M. Z., W. M. and X.-P. X. fabricated the CPLN and PPLN devices, supervised by Q. Z..
L. G. and C. S. fabricated the Si$_3$N$_4$ devices, supervised by J. Liu.
B. S., Y. Z., M. Z. and J. Liu analyzed the data and prepared the manuscript with input from others. 
J. Liu and Q. Z. managed the collaboration.
J. Liu initiated and supervised the project. 

\noindent \textbf{Conflict of interest}:
B. S., Y.-H. L., J. Long and J. Liu filed a patent application related to this work.  
Others declare no conflicts of interest.

\noindent \textbf{Data Availability Statement}: 
The code and data used to produce the plots within this work will be released on the repository \texttt{Zenodo} upon publication of this preprint.

\end{footnotesize}
\bibliographystyle{apsrev4-1}
\bibliography{bibliography}
\end{document}

\title{Supplementary Materials for: An alkali-referenced vector spectrum analyzer for visible-light integrated photonics}

\author{Baoqi Shi}
\affiliation{International Quantum Academy, Shenzhen 518048, China}
\affiliation{Department of Optics and Optical Engineering, University of Science and Technology of China, Hefei 230026, China}

\author{Ming-Yang Zheng}
\affiliation{Jinan Institute of Quantum Technology and CAS Center for Excellence in Quantum Information and Quantum Physics, University of Science and Technology of China, Jinan 250101, China}
\affiliation{Hefei National Laboratory, University of Science and Technology of China, Hefei 230088, China}

\author{Yunkai Zhao}
\affiliation{International Quantum Academy, Shenzhen 518048, China}
\affiliation{Shenzhen Institute for Quantum Science and Engineering, Southern University of Science and Technology, Shenzhen 518055, China}

\author{Yi-Han Luo}
\affiliation{International Quantum Academy, Shenzhen 518048, China}

\author{Jinbao Long}
\affiliation{International Quantum Academy, Shenzhen 518048, China}

\author{Wei Sun}
\affiliation{International Quantum Academy, Shenzhen 518048, China}

\author{Wenbo Ma}
\affiliation{Jinan Institute of Quantum Technology and CAS Center for Excellence in Quantum Information and Quantum Physics, University of Science and Technology of China, Jinan 250101, China}

\author{Xiu-Ping Xie}
\affiliation{Jinan Institute of Quantum Technology and CAS Center for Excellence in Quantum Information and Quantum Physics, University of Science and Technology of China, Jinan 250101, China}
\affiliation{Hefei National Laboratory, University of Science and Technology of China, Hefei 230088, China}

\author{Lan Gao}
\affiliation{International Quantum Academy, Shenzhen 518048, China}
\affiliation{Shenzhen Institute for Quantum Science and Engineering, Southern University of Science and Technology, Shenzhen 518055, China}

\author{Chen Shen}
\affiliation{International Quantum Academy, Shenzhen 518048, China}
\affiliation{Qaleido Photonics, Shenzhen 518048, China}

\author{Anting Wang}
\affiliation{Department of Optics and Optical Engineering, University of Science and Technology of China, Hefei 230026, China}

\author{Wei Liang}
\affiliation{Suzhou Institute of Nano-tech and Nano-bionics, Chinese Academy of Sciences, Suzhou 215123, China}

\author{Qiang Zhang}
\affiliation{Jinan Institute of Quantum Technology and CAS Center for Excellence in Quantum Information and Quantum Physics, University of Science and Technology of China, Jinan 250101, China}
\affiliation{Hefei National Laboratory, University of Science and Technology of China, Hefei 230088, China}
\affiliation{Hefei National Research Center for Physical Sciences at the Microscale and School of Physical Sciences, University of Science and Technology of China, Hefei 230026, China}
\affiliation{CAS Center for Excellence in Quantum Information and Quantum Physics, University of Science and Technology of China, Hefei 230026, China}

\author{Junqiu Liu}
\affiliation{International Quantum Academy, Shenzhen 518048, China}
\affiliation{Hefei National Laboratory, University of Science and Technology of China, Hefei 230088, China}

\maketitle

%%%%%%%%%%%%%%%%%%%%%%%%%%
%%%%%%%%%%%%%%%%%%%%%%%%%%
%%%%%%%%%%%%%%%%%%%%%%%%%%
\section{Fiber cavity characterization}
\vspace{0.5cm}

Since both the NIR and SH chirping lasers are frequency-calibrated using the NIR transmission spectrum of the fiber cavity, we calibrate the fiber cavity's FSR using a sideband modulation method described in Ref. \cite{Luo:24}. 
This method requires three neighboring resonances to measure the local FSR of the fiber cavity at different wavelength, and can account for dispersion-induced FSR variation over the measurement bandwidth. 
We measure the fiber cavity's FSR from 182.8 to 237.9 THz (55.1 THz) and fit the data with a cubic polynomial formula, as shown in Supplementary Fig. \ref{SIfig:fiberloop}a.
This wide measurement range allows calibration of the fiber cavity's dispersion up to the third order.
The fit residuals are displayed in Supplementary Fig. \ref{SIfig:fiberloop}b, showing a standard deviation of 368 Hz.

%%%%%%%%%%%%%%%%%%%%%%%%%%%%%%
\begin{figure}[h!]
\centering
\renewcommand{\figurename}{Supplementary Figure}
\includegraphics{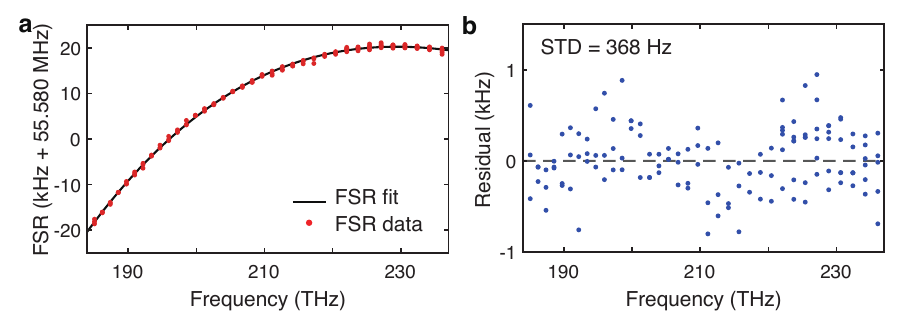}
\caption{
\textbf{Calibration of the fiber cavity}. 
\textbf{a}. 
Measured fiber cavity's FSR variation from 182.8 to 237.9 THz.
The FSR curve is fitted with a cubic polynomial formula.
\textbf{b}. 
The fit residual has a standard deviation (STD) of 368 Hz.
}
\label{SIfig:fiberloop}
\end{figure}
%%%%%%%%%%%%%%%%%%%%%%%%%%%%%%

%\clearpage

%%%%%%%%%%%%%%%%%%%%%%%%%%
\section{Design and fabrication of CPLN and PPLN waveguides}
\vspace{0.5cm}

\begin{figure}[h!]
\centering
\renewcommand{\figurename}{Supplementary Figure}
\includegraphics[width=\linewidth]{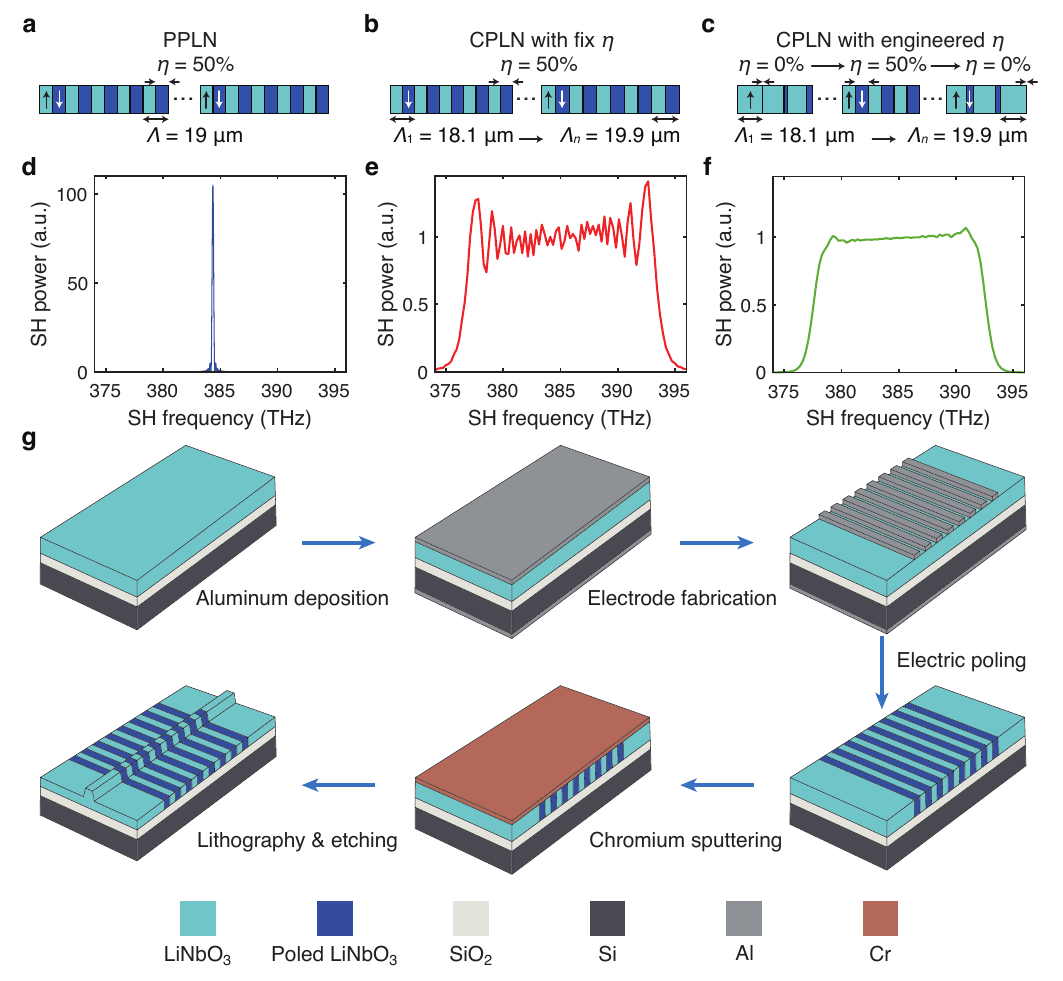}
\caption{
\textbf{Design and fabrication of CPLN and PPLN waveguides}. 
\textbf{a-c}. 
Illustration of poling period $\Lambda$ and duty cycle $\eta$ of a PPLN waveguide, a CPLN waveguide of fixed $\eta=50\%$, and a CPLN waveguide of engineered $\eta$.
\textbf{d-f}. 
Theoretical SH power curves generated from the three waveguides.
\textbf{g}. 
Fabrication process flow of CPLN waveguides.
}
\label{SIfig:CPLN}
\end{figure}

In the regime of low conversion efficiency, the SHG efficiency of a PPLN waveguide is given by
\begin{equation}
\left|\widehat{D}(\omega)\right|^2= \left|\frac{2 \pi}{\lambda_{\mathrm{FH}} n_{\mathrm{SH}}} \int_{-\infty}^{\infty}\mathrm{d} z~\tilde{d}(z) \exp \left(-i \Delta k^{\prime} z\right) \right|^2,
\label{Eq.D}
\end{equation}
where, $\Delta k'=k'_\text{SH}-2k'_\text{FH}$ is the phase velocity mismatch between the FH (fundamental harmonic) and SH (second harmonic) waves, 
$n_{\mathrm{SH}}$ is the SH wave's effective refractive index, 
$\lambda_{\mathrm{FH}}$ is the FH wave's wavelength.
In Eq. \ref{Eq.D}, $\tilde{d}(z)$ is the nonlinear optical coefficient along the propagation direction $z$, and is expressed as
\begin{equation}
\tilde{d}(z)=\mathrm{rect}(z) \frac{2}{\pi}\sin[\pi \eta(z)]\exp\left[-i K_\mathrm{QPM}(z)z\right].
\label{Eq.d}
\end{equation}
where $\eta(z)$ is the poling duty cycle, 
$K_\mathrm{QPM}=2\pi/\Lambda(z)$ with $\Lambda(z)$ being the poling period. 
In our case, the group velocity mismatch is ignored so that $\widehat{D}(\omega)$ is the exact Fourier transform of $\tilde{d}(z)$ along the propagation direction\cite{Imeshev:00}.
Supplementary Fig. \ref{SIfig:CPLN}(a-c) show three designs: a PPLN waveguide, a CPLN waveguide of fixed duty cycle $\eta=50\%$, and a CPLN waveguide of varying duty cycle.
Supplementary Fig. \ref{SIfig:CPLN}(d-f) compares the theoretical SH power generated from these three designs with the same input power.
All these three waveguides are 20 mm long.

The PPLN waveguide has a uniform poling period of $\Lambda=19$ $\mu$m and $\eta=50\%$.
Its theoretical output SH power curve reveals that the SH bandwidth is limited to 138 GHz.
A varying poling period $\Lambda$ can extend the SH bandwidth.
For the CPLN waveguide with $\eta=50\%$ duty cycle and varying $\Lambda$ from 18.1 $\mu$m to 19.9 $\mu$m along the propagating direction, the theoretical output SH power curve indicates a SH bandwidth reaching 16.3 THz.
However, a 20\% power fluctuation is accompanied, introduced by the abrupt changes in $\tilde{d}(z)$ at the two ends of the CPLN waveguide.

The broadband fluctuation of SH power can be ameliorated through poling duty cycle engineering.
Supplementary Fig. \ref{SIfig:CPLN}c shows the CPLN waveguide with increasing $\eta$ from 0$\%$ to 50$\%$ over the first 3 mm waveguide length, and decreasing $\eta$ from 50$\%$ to 0$\%$ over the last 3 mm. 
The duty cycle remains at $\eta=50\%$ in the middle section of the waveguide.
Consequently, the SH power's fluctuation is reduced, as shown in Supplementary Fig. \ref{SIfig:CPLN}f.
Meanwhile, an SH bandwidth of 14.1 THz is achieved.

The fabrication process flow of CPLN waveguide is shown in Supplementary Fig. \ref{SIfig:CPLN}g.
The waveguide is fabricated on a 10-um-thick z-cut MgO-doped LiNbO$_3$ film bonded onto a 2-µm-thick thermal wet SiO$_2$ wafer (NANOLN Inc.). 
First, aluminum is deposited on both sides of the wafer by electron-beam evaporation.
Metal electrodes are fabricated using standard UV lithography and wet etching.
Next, an external electric field is applied to realize periodic poling of the LiNbO$_3$ film. 
Afterward, chromium is sputtered on the LiNbO$_3$ film, which serves as the etching mask to etch LiNbO$_3$ ridge waveguides. 
Finally, inductively coupled plasma (ICP) etching is used to transfer the waveguide pattern from the chromium mask to the LiNbO$_3$ film and form the CPLN waveguides. 

%%%%%%%%%%%%%%%%%%%%%%%%%%%%%%
%%%%%%%%%%%%%%%%%%%%%%%%%%%%%%

\clearpage

%%%%%%%%%%%%%%%%%%%%%%%%%%
\section{Characterization of an air-gap Fabry–Pérot cavity}
\vspace{0.5cm}

To evaluate our VSA's practical bandwidth around 780 nm wavelength, we characterize an air-gap Fabry–Pérot (FP) cavity's transmission spectrum.
As shown in Supplementary Fig. \ref{SIfig:FP}a, the FP cavity is formed by attaching a planar mirror (5 mm diameter, 1 mm thickness) and a concave mirror (5 mm diameter, 2.3 mm thickness, -250 mm radius of curvature) to a 7.16-mm-long hollow fused silica spacer. 
Supplementary Fig. \ref{SIfig:FP}b shows a zoom-in of the FP cavity's transmission spectrum with resonances marked by red dots.
The resonance's signal-to-noise ratio (SNR) is calculated using $\mathrm{SNR}=P_\mathrm{peak}/P_\mathrm{noise}$, where $P_\mathrm{peak}$ is resonance peak's power, and $P_\mathrm{noise}$ is the standard deviation of the noise power near the resonance.
Supplementary Fig. \ref{SIfig:FP}c shows the calculated resonance SNR within our VSA bandwidth from 766 to 795 nm. 
Every resonance has an SNR exceeding 22.3 dB, demonstrating that our VSA has a practical bandwidth of 14.3 THz.
Additionally, we can measure the loaded $Q$ factor of the FP cavity.
Supplementary Fig. \ref{SIfig:FP}d shows a resonance with a Lorentzian fit. 
The resonance exhibits 86.3 MHz linewidth, corresponding to $Q=4.48\times10^6$.

With alkali-referenced absolute frequency calibration, our VSA allows measurement of the frequency drift of each resonance of the FP cavity. 
We repeat the measurement of the FP cavity's transmission spectrum ten times over 120 seconds, and calculate the frequency drift for each resonance. 
The average frequency drift for each measurement is shown in Supplementary Fig. \ref{SIfig:FP}e. 
The resonances exhibit a drift of 7.6 MHz in 120 s.
Since the FP cavity's mirrors are attached to the fused silica spacer which has a thermal expansion coefficient of $5.5\times10^{-7}$/K, we can determine the temperature deviation of the FP cavity to be -36 mK.

%%%%%%%%%%%%%%%%%%%%%%%%%%%%%%
\begin{figure}[h!]
\centering
\renewcommand{\figurename}{Supplementary Figure}
\includegraphics[width=\linewidth]{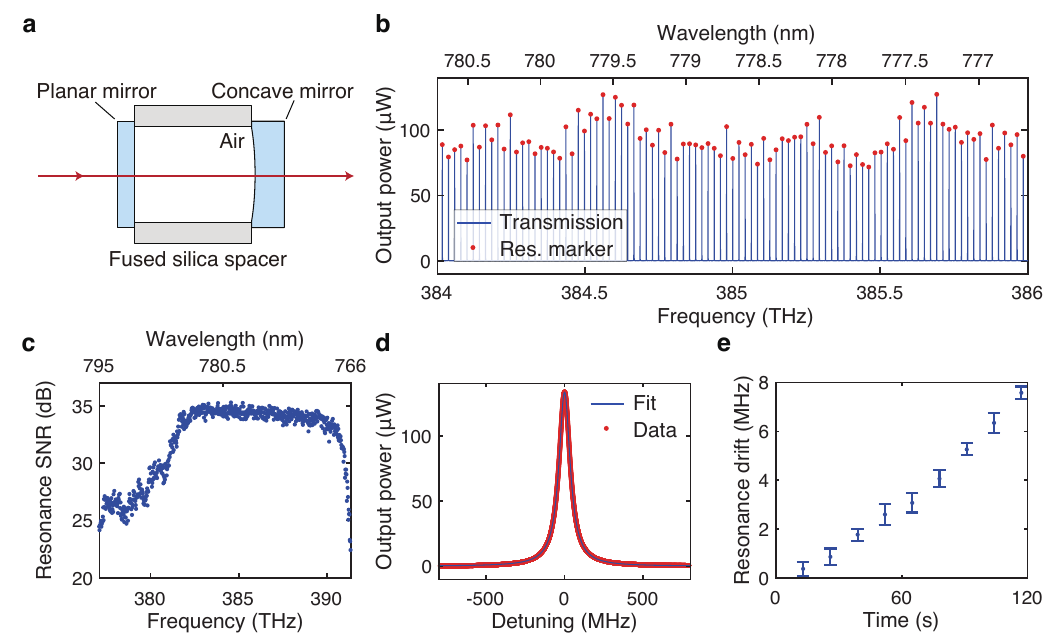}
\caption{
\textbf{Characterization of an air-gap Fabry–Pérot cavity}.
\textbf{a}. 
Illustration of the FP cavity's structure.
\textbf{b}.
Zoom-in of the FP cavity's transmission spectrum measured by our VSA.
Resonances are marked by red dots.
\textbf{c}.
Measured signal-to-noise ratio (SNR) of resonances within our VSA bandwidth.
All resonances have SNR over 22.3 dB.
\textbf{d}.
Measured transmission of one resonance.
The resonance has a fitted linewidth of 86.3 MHz, corresponding to a loaded quality factor $Q=4.48\times10^6$.
\textbf{e}.
Measured resonance drift in 120 s.
The 7.6 MHz frequency drift corresponds to -36 mK temperature change of the FP cavity.
Error bar is the standard deviation of 680 resonances' frequency drift.
}
\label{SIfig:FP}
\end{figure}
%%%%%%%%%%%%%%%%%%%%%%%%%%%%%%

\clearpage

%%%%%%%%%%%%%%%%%%%%%%%%%%
\section{The second-harmonic laser's dynamic linewidth}
\vspace{0.5cm}

\begin{figure}[h!]
\centering
\renewcommand{\figurename}{Supplementary Figure}
\includegraphics[width=\linewidth]{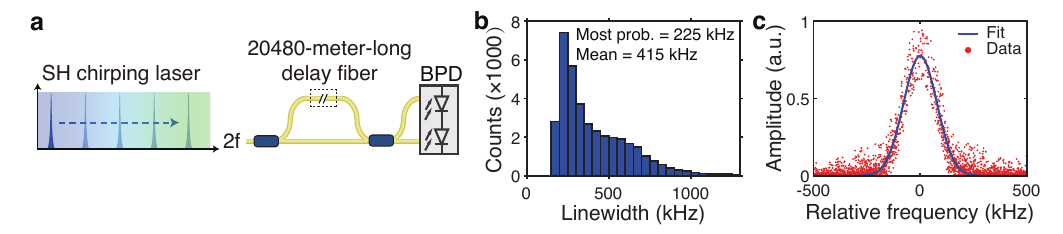}
\caption{
\textbf{Characterization of the SH laser's dynamic linewidth}. 
\textbf{a}. 
Schematic of the self-delayed heterodyne measurement. 
\textbf{b}. 
Histogram of measured SH chirping laser's dynamic linewidth. 
Here 37,945 segments of beatnote signals are taken and processed with FFT to extract the laser linewidth. 
\textbf{c}. 
Frequency-domain data of the beat signal with linewidth around 225 kHz.
}
\label{SIfig:linewidth}
\end{figure}

Supplementary Fig. \ref{SIfig:linewidth}a shows the self-delayed heterodyne setup to measure the SH laser's dynamic linewidth. 
The SH laser is split into two paths. 
One path travels through a 20,408-meter-long fiber, introducing a time delay exceeding the laser's coherent time. 
This delayed path is subsequently recombined with the other undelayed path.
During the measurement, the SH laser chirps at 12.5 THz/s rate. 
The temporal beat signal is directly recorded and then converted into the frequency domain using short-time Fourier transform (STFT) with 100-$\mu$s window length. 

We analyze the linewidth of 37,948 beatnote signals. 
The result histogram is shown in Supplementary Fig. \ref{SIfig:linewidth}b. 
The statistic laser linewidth corresponding to 100-$\mu$s window is determined to be 415 kHz, and the most probable value is 225 kHz.
The frequency-domain data of the beat signal with linewidth around the most probable value is shown in Supplementary Fig. \ref{SIfig:linewidth}c.
This measured laser linewidth is determined by the laser's intrinsic linewidth, chirp nonlinearity, as well as the fiber delay's fluctuation.
Overall, we conclude that the frequency resolution of our VSA is 415 kHz.

%%%%%%%%%%%%%%%%%%%%%%%%%%%%%%
%%%%%%%%%%%%%%%%%%%%%%%%%%%%%%

\clearpage

%%%%%%%%%%%%%%%%%%%%%%%%%%
\section{Fabrication process of silicon nitride integrated waveguide}
\vspace{0.5cm}

The Si$_3$N$_4$ integrated waveguide is fabricated using an optimized deep-ultraviolet (DUV) subtractive process on 6-inch wafers \cite{Ye:23}. 
The process flow is illustrated in Supplementary Fig. \ref{SIfig:SiN}. 
First, 300-nm-thick Si$_3$N$_4$ film is deposited on a clean thermal wet SiO$_2$ substrate via low-pressure chemical vapor deposition (LPCVD). 
Subsequently, an SiO$_2$ film is deposited on Si$_3$N$_4$ as an etch hardmask, again via LPCVD. 
Next, DUV stepper lithography is performed, followed by dry etching to transfer the pattern from the photoresist to the SiO$_2$ hardmask, and then to the Si$_3$N$_4$ layer. 
The dry etching uses etchants comprising CHF$_3$ and O$_2$ to create ultra-smooth and vertical etched sidewalls, critical for minimizing optical losses in waveguides. 
Afterward, the photoresist is removed, and thermal annealing of the entire waver is applied under nitrogen atmosphere at 1200 $^\circ$C. 
The annealing eliminates hydrogen contents, which can contribute to optical absorption losses in the waveguides.
Then a 3-$\mu$m-thick SiO$_2$ top cladding layer is deposited on the wafer, followed by another thermal annealing at 1200 $^\circ$C to remove residual hydrogen content.
Finally, UV photolithography and deep dry etching are performed to create smooth chip facets and to define the chip size. The wafer is separated into individual chips through backside grinding or dicing.

%%%%%%%%%%%%%%%%%%%%%%%%%%%%%%
\begin{figure}[h!]
\centering
\renewcommand{\figurename}{Supplementary Figure}
\includegraphics[width=\linewidth]{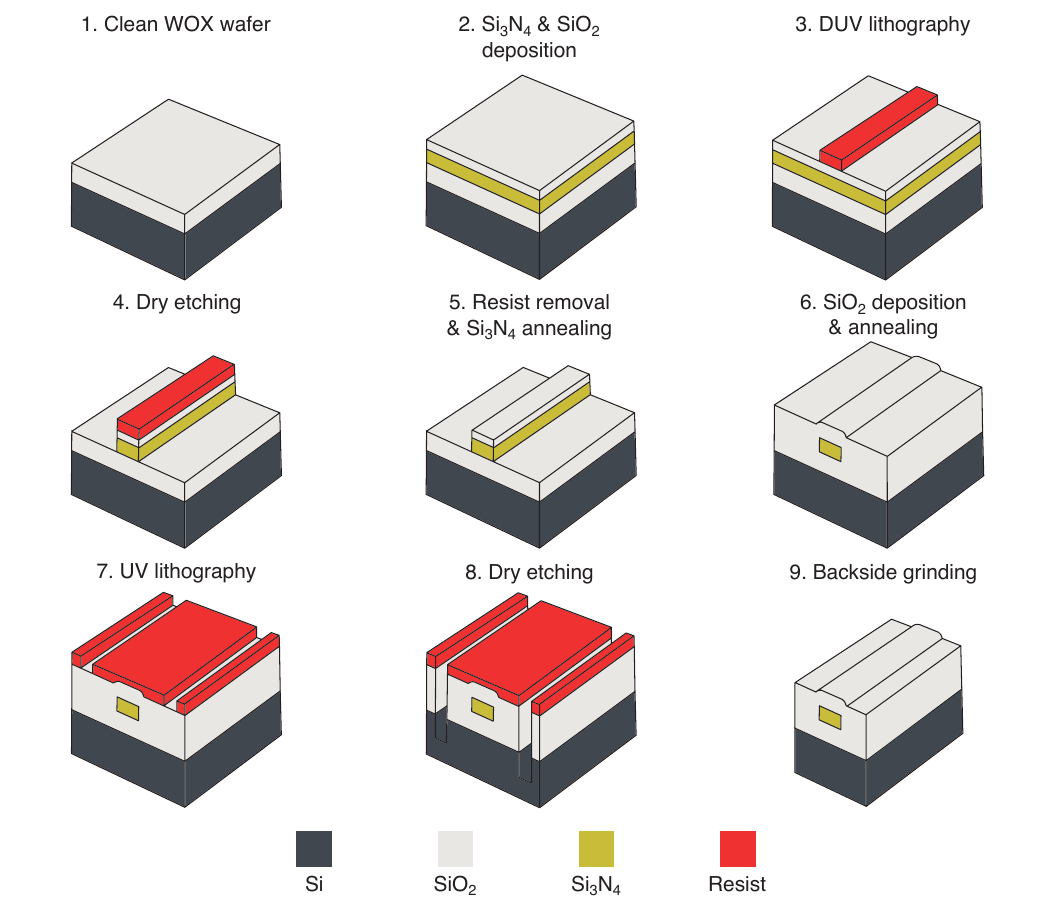}
\caption{
\textbf{The DUV subtractive process flow of 6-inch-wafer Si$_3$N$_4$ foundry fabrication}. 
WOX, thermal wet oxide (SiO$_2$).
}
\label{SIfig:SiN}
\end{figure}
%%%%%%%%%%%%%%%%%%%%%%%%%%%%%%

\clearpage

%%%%%%%%%%%%%%%%%%%%%%%%%%
\section{Inverse taper for efficient fiber-chip coupling}
\vspace{0.5cm}

Central to applications of integrated photonics is the capability to efficiently couple light into and out of chips via optical fibers. 
Here, we employ inverse waveguide tapers \cite{Almeida:03, Ren:11} on the chip facets to achieve mode matching between the Si$_3$N$_4$ waveguide mode and the lensed fiber mode, as illustrated in Supplementary Fig. \ref{SIfig:taper}a. 
Experimentally, we use a lensed fiber with 2.5-$\mu$m-diameter focus spot, the smallest size commercially available.
To optimize coupling efficiency, we design, fabricate, and measure waveguides with varying taper widths ($w$) to match the lensed fiber mode. 
At 780 nm wavelength, for both TE and TM polarization, the measured coupling efficiency (for two chip facets, in and out) of different taper widths $w$ is shown in Supplementary Fig. \ref{SIfig:taper}b.
The result indicates that, with $w=150$ nm taper width (the smallest dimension feasible with our DUV stepper lithography), the inverse waveguide taper achieves the highest coupling efficiency, with 40\% for the TE and 43\% for the TM polarization. 

%%%%%%%%%%%%%%%%%%%%%%%%%%%%%%
\begin{figure}[h!]
\centering
\renewcommand{\figurename}{Supplementary Figure}
\includegraphics{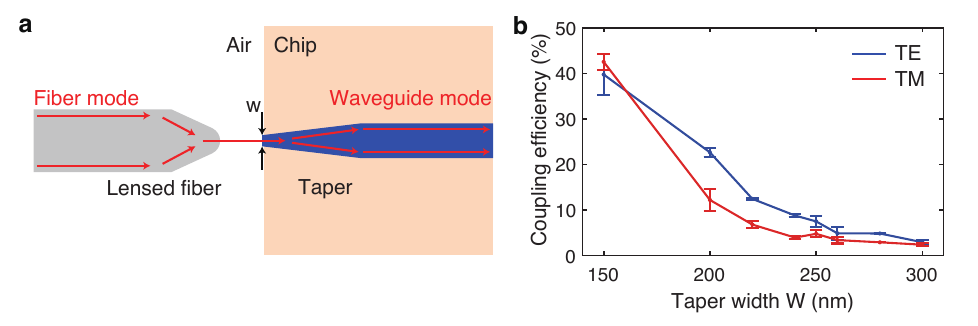}
\caption{
\textbf{Schematic and characterization of inverse waveguide tapers at 780 nm wavelength}. 
\textbf{a}. 
Schematic of an inverse taper to facilitate mode matching and to achieve efficient light coupling from a lensed fiber into the chip.
\textbf{b}. 
Measured coupling efficiency (including two chip facets) versus different taper widths $w$, for both TE and TM polarization at 780 nm wavelength.
}
\label{SIfig:taper}
\end{figure}
%%%%%%%%%%%%%%%%%%%%%%%%%%%%%%

%%%%%%%%%%%%%%%%%%%%%%%%%%
\section{Polarization purification}
\vspace{0.5cm}

In integrated waveguides, TE and TM optical modes experience different optical properties of the waveguide, such as group index, loss, and coupling strength.
To avoid coupling to undesired polarization and more accurately evaluate resonance contrast, it is critical to ensure broadband polarization purity of the input laser. 
Even though polarization-maintaining (PM) fibers are used, the laser coupled into the chip can have degraded polarization purity due to various optical components and fiber connectors it passes. 
The fiber birefringence can transform a linearly polarized laser into complex polarization states at different wavelengths.
Therefore, it is impossible to correct the polarization simply using a polarization controller (PC).
We experimentally examine the laser polarization before and after a PC, using a polarization analyzer. 
The PC is adjusted to optimize the polarization extinction ratio (PER) and polarization direction at 1550 nm.
The laser PER and the deviation from TE polarization at different wavelengths are shown in Supplementary Fig. \ref{SIfig:pol}b.
The PC can only correct the laser polarization within a limited bandwidth around 1550 nm. 
For a wider range, the polarization correction made by the PC is insufficient due to different polarization at different wavelength.

To purify laser polarization, we employ a polarizer unit as shown in Supplementary Fig. \ref{SIfig:pol}a.
The polarizer unit contains a pair of coupled collimators.
Between the collimators, a broadband polarizer is inserted to filter out light of undesired polarization.
In our VSA, we use PM lensed fibers for light coupling with high polarization stability.
To align the polarization with the slow axes of the PM fiber connected to the receiver coupler, we use a broadband half-wave plate for adjustment.
The polarization purification performance of the polarizer unit is shown in Supplementary Fig. \ref{SIfig:pol}c.
With this unit, an average PER of 37 dB over 160 nm bandwidth is achieved, showing 7 dB enhancement compared to the case without the unit.
Additionally, the mean deviation degree is optimized from 14.8$^\circ$ to 1.1$^\circ$.

We further characterize laser polarization degradation after passing through a fiber connector.
First, we purify the laser polarization using our polarizer unit, and couple the polarization-purified laser into the slow axes of a PM fiber.
Supplementary Fig. \ref{SIfig:pol}d shows the analyzed laser polarization in the PM fiber is analyzed (red dots).
Next, the PM fiber is connected to another PM fiber using a fiber connector, and the output laser polarization is analyzed again, as shown by the black dots in Supplementary Fig. \ref{SIfig:pol}d.
The average PER of the laser is degraded by 11 dB after passing through the fiber connector.
The polarization direction rotates to 9$^\circ$ at some wavelength.
This degradation originates from the misalignment of the slow axes of the two PM fibers.
It evidences the necessity to place the polarizer unit right before the lensed fiber, to avoid any disturbance from fiber connectors. 

To demonstrate the importance of polarization purification, we characterize a 120-GHz-FSR Si$_3$N$_4$ microresonator with and without purification.
Supplementary Fig. \ref{SIfig:pol}e shows the microresonator transmission spectrum. 
Without purification, the TM mode resonances (marked by green dots) are seen, which affects correct evaluation of TE resonances' extinction ratio (marked by red dots).
With purification, all TM resonances vanish, and TE resonances can now achieve critical coupling (zero transmittance).

Polarization purity influences the resonance fit and renders inaccurate extrapolation of loss values. 
As shown in Supplementary Fig. \ref{SIfig:pol}f, without purification, the fitted loss values are $\kappa_0/2\pi=22.4$ MHz and $\kappa_\mathrm{ex}/2\pi=8.2$ MHz. 
With purification, the fitted loss values are $\kappa_0/2\pi=15.7$ MHz and $\kappa_\mathrm{ex}/2\pi=15.7$ MHz.

%%%%%%%%%%%%%%%%%%%%%%%%%%%%%%
\begin{figure}[h!]
\centering
\renewcommand{\figurename}{Supplementary Figure}
\includegraphics[width=\linewidth]{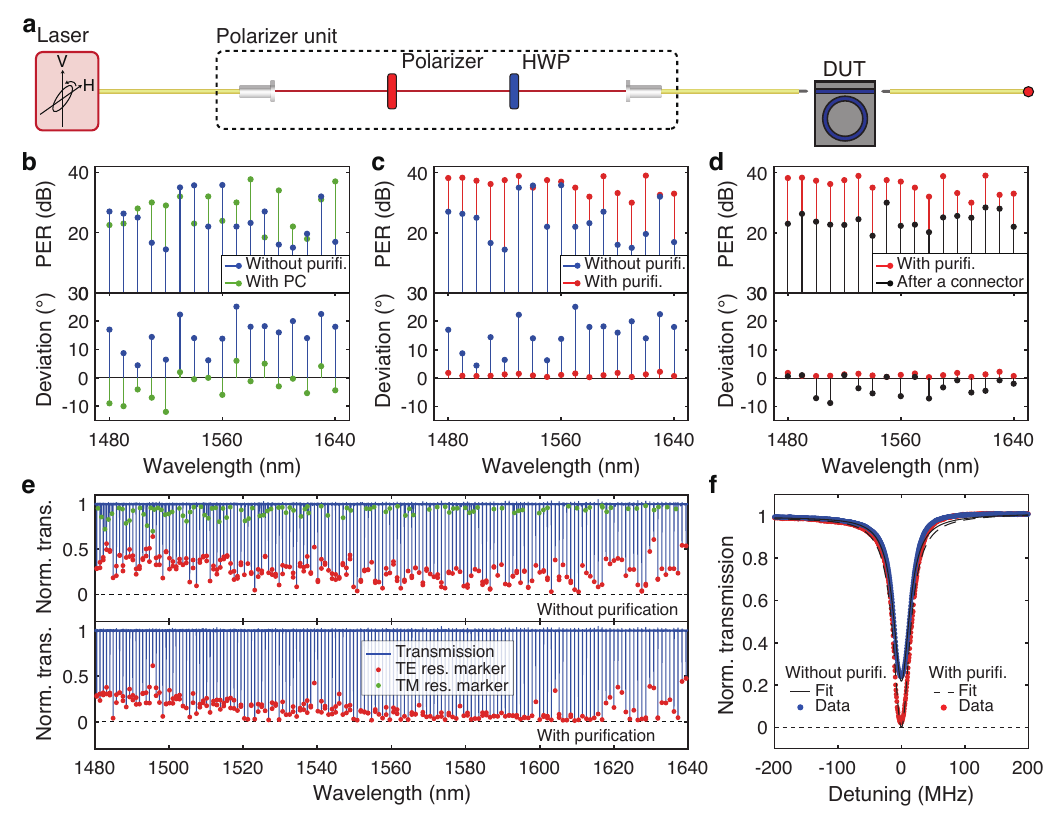}
\caption{
\textbf{Polarization purification setup}.
\textbf{a}. 
Schematic and setup of polarization purification.
HWP, half-wave plate.
\textbf{b}.
Comparison of laser polarization adjusted by PC and without purification.
The top panel shows the polarization extinction ratio (PER).
The bottom panel shows the deviation degree from the TE polarization.
PC, polarization controller.
\textbf{c}.
Comparison of laser polarization with and without purification.
\textbf{d}.
Comparison of the polarization of the purified laser and the laser passing through a fiber connector.
\textbf{e}.
Transmission spectra of a 120-GHz-FSR Si$_3$N$_4$ microresonator without (top) and with (bottom) polarization purification.
\textbf{f}.
An example of fitted resonance profile.
Without purification, the fitted loss values are $\kappa_0/2\pi=22.4$ MHz and $\kappa_\text{ex}/2\pi=8.2$ MHz.
With purification, the fitted loss values are $\kappa_0/2\pi=15.7$ MHz and $\kappa_\text{ex}/2\pi=15.7$ MHz.
}
\label{SIfig:pol}
\end{figure}
%%%%%%%%%%%%%%%%%%%%%%%%%%%%%%

%\clearpage

%%%%%%%%%%%%%%%%%%%%%%%%%%
\section{Vector spectrum analyzer in the 1064 nm wavelength}
\vspace{0.5cm}

We build a separate VSA operated in the 1064 nm wavelength as shown in Supplementary Fig. \ref{SIfig:1um}a.
An ECDL(Toptica CTL) is used, whose mode-hop-free chirp range spans from 1020.6 to 1096.6 nm (293.7 to 273.4 THz), covering 20.3 THz bandwidth. 
To calibrate the laser's relative frequency, we use a fiber cavity made of single-mode fibers for 1064 nm wavelength.
We calibrate the fiber cavity's FSR using a sideband modulation method illustrated in Ref. \cite{Luo:24}. 
The fiber cavity's FSR is measured across the ECDL bandwidth and fitted by a quadratic polynomial formula, as shown in Supplementary Fig. \ref{SIfig:1um}c.
The absolute frequency is referenced to the ECDL's built-in wavelength trigger.

The ECDL passes through the polarizer unit and is coupled into the DUT (Si$_3$N$_4$ microresonator) to characterize microresonator loss and dispersion.
Supplementary Fig. \ref{SIfig:1um}b top panel shows the microresonator's transmission spectrum.
Via peak searching, each resonance is identified and marked by red dots.
A typical resonance, along with its fit, is displayed in Supplementary Fig. \ref{SIfig:1um}d.
The fitted loss values are $\kappa_0/2\pi=28.1$ MHz and $\kappa_\mathrm{ex}/2\pi=14.3$ MHz.
Within our 20.3 THz bandwidth, we can find, measure and fit 396 resonances and the results are shown in Fig. \ref{SIfig:1um}b bottom panel.
The $\kappa_0/2\pi$ histogram is shown in Supplementary Fig. \ref{SIfig:1um}e.
The most probable value of $\kappa_0/2\pi=33$ MHz corresponds to an intrinsic quality factor of $Q_0=8.6 \times 10^6$.

%%%%%%%%%%%%%%%%%%%%%%%%%%%%%%
\begin{figure}[h!]
\centering
\renewcommand{\figurename}{Supplementary Figure}
\includegraphics[width=\linewidth]{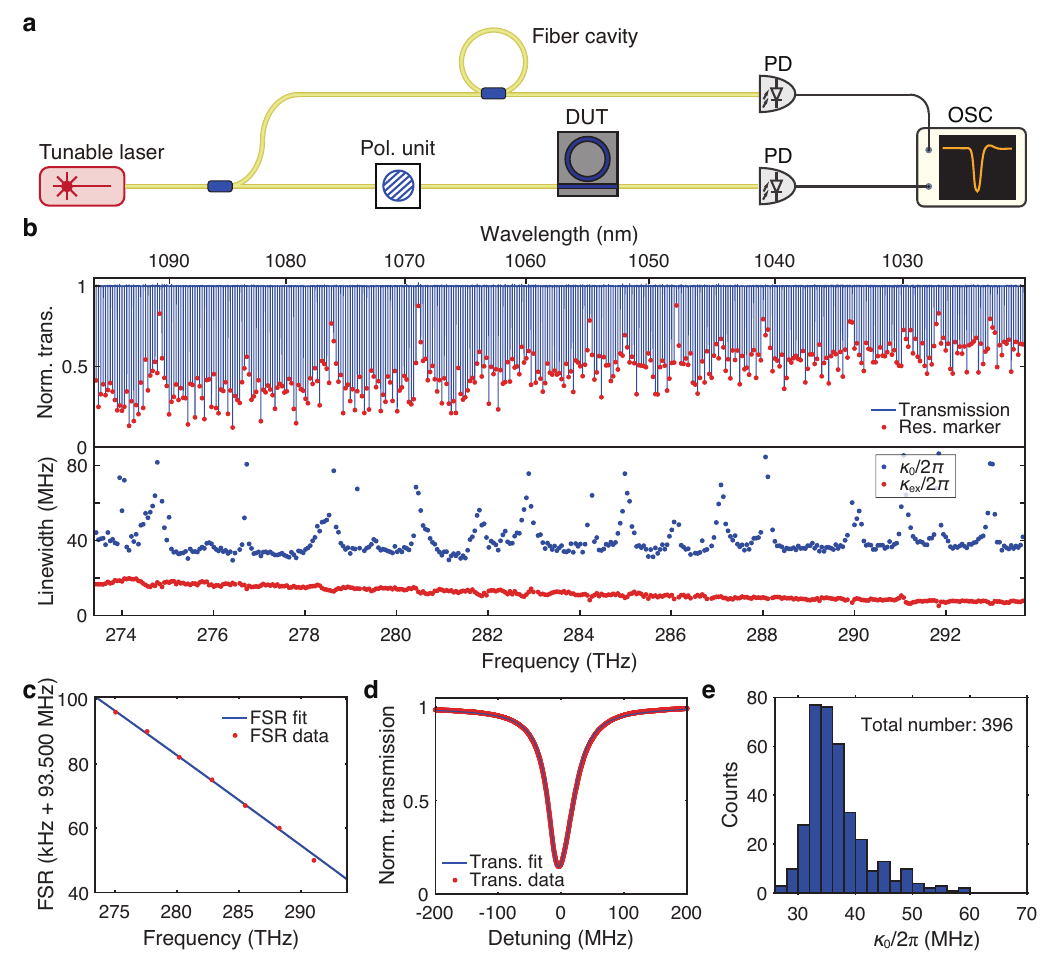}
\caption{
\textbf{Characterization of the Si$_3$N$_4$ microresonator from 1020.6 to 1096.6 nm wavelength}.
\textbf{a}. 
Experimental setup to characterize the transmission (loss) and dispersion of Si$_3$N$_4$ microresonators in this wavelength range. 
\textbf{b}.
Measured microresonator transmission spectrum (top) and resonance linewidth (bottom).
\textbf{c}.
Measured fiber cavity's FSR variation.
The FSR curve is fitted with a quadratic polynomial formula.
\textbf{d}.
A typical measured resonance profile.
The resonance has a fitted $\kappa_0/2\pi=28.1$ MHz and $\kappa_{ex}/2\pi=14.3$ MHz.
\textbf{e}.
Histogram of 396 resonances from the microresonator, showing the most probable value of $\kappa_0/2\pi=33$ MHz.
}
\label{SIfig:1um}
\end{figure}
%%%%%%%%%%%%%%%%%%%%%%%%%%%%%%

\newpage

%%%%%%%%%%%%%%%%%%%%%%%%%%%%%%%%%%%%%%%%%%%%%%%%%%%%%%%%
\section{OFDR setup and principle}
\vspace{0.5cm}

We configure our VSA as an optical frequency-domain reflectometry (OFDR) to characterize integrated long spiral waveguides, as illustrated in Supplementary Fig. \ref{SIfig:OFDR}. 
A circulator is connected to Port 1.
Through the circulator, the SH laser is coupled into the DUT. 
The reflected light from the DUT (spiral waveguide) is connected to Port 3 and probed by the photodetector.

The reflected light at a distance $l$ within the waveguide of physical length $L_0$ has an amplitude $A_l$. 
The strong reflection signal from the waveguide's front facet (i.e. $l=0$) maintains a constant amplitude $A_0$. 
These two fields, $A_l$ and $A_0$, interfere and produce a beat signal characterized by frequency $\Delta f=2\gamma n_g l/c$, where $c$ is the speed of light in air, $\gamma$ denotes the laser chirp rate, and $n_g$ is group index. 
Therefore, the photodetected beat signal can be expressed as $V_l(t)=A_0 A_l \cos(2\pi\Delta ft)$.
The laser chirp rate $\gamma$ is determined through the relative frequency calibration with the fiber cavity.
With FFT, the relative reflection power from different location within the waveguide is calculated.

%%%%%%%%%%%%%%%%%%%%%%%%%%%%%%
\begin{figure}[h!]
\centering
\renewcommand{\figurename}{Supplementary Figure}
\includegraphics[width=\linewidth]{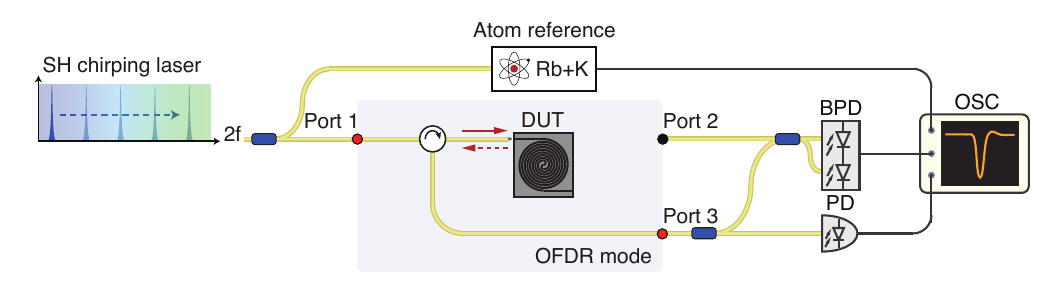}
\caption{
\textbf{Experimental setup of OFDR}. 
}
\label{SIfig:OFDR}
\end{figure}
%%%%%%%%%%%%%%%%%%%%%%%%%%%%%%

\newpage

%%%%%%%%%%%%%%%%%%%%%%%%%%%%%%%%%%%%%%%%%%%%%%%%%%%%%%%%
\section{Mode-locked laser spectrum}
\vspace{0.5cm}

The experimental setup to characterize mode-locked laser spectra is shown in Supplementary Fig. \ref{SIfig:BPF}a.
The Port 1 and 2 are shortcut, and the mode-locked laser is connected to Port 3.
The chirping SH laser beats against each CW component of the mode-locked laser, and the beat signal is detected by the balanced detector and recorded by the oscilloscope.
The beat signal is digitally processed, as the flowcharts shown in Supplementary Fig. \ref{SIfig:BPF}(b-e).
Supplementary Fig. \ref{SIfig:BPF}b shows the time-domain beat signal when the SH laser chirps across a CW component of the mode-locked laser.
The beat signal is processed by a digitally implemented finite impulse response (FIR) band-pass filter (BPF).
The BPF has a 10 MHz passband center frequency and a 6 MHz bandwidth.
Supplementary Fig. \ref{SIfig:BPF}c shows filtered beat signal. 
Then Hilbert transforms \cite{Marple:99} is applied to obtain the pulse's envelope, as shown in Supplementary Fig. \ref{SIfig:BPF}d.
This operation creates a pair of pulses for each CW component, whose center frequency corresponds to the frequency of the CW component.
The two pulses in a pair have the same shape but 20 MHz frequency difference.
Via deconvolution, one pulse remains and is frequency-translated to the center of the pulse pair, i.e. the frequency the CW component, as shown in Supplementary Fig. \ref{SIfig:BPF}e.

%%%%%%%%%%%%%%%%%%%%%%%%%%%%%%
\begin{figure}[h!]
\centering
\renewcommand{\figurename}{Supplementary Figure}
\includegraphics[width=\linewidth]{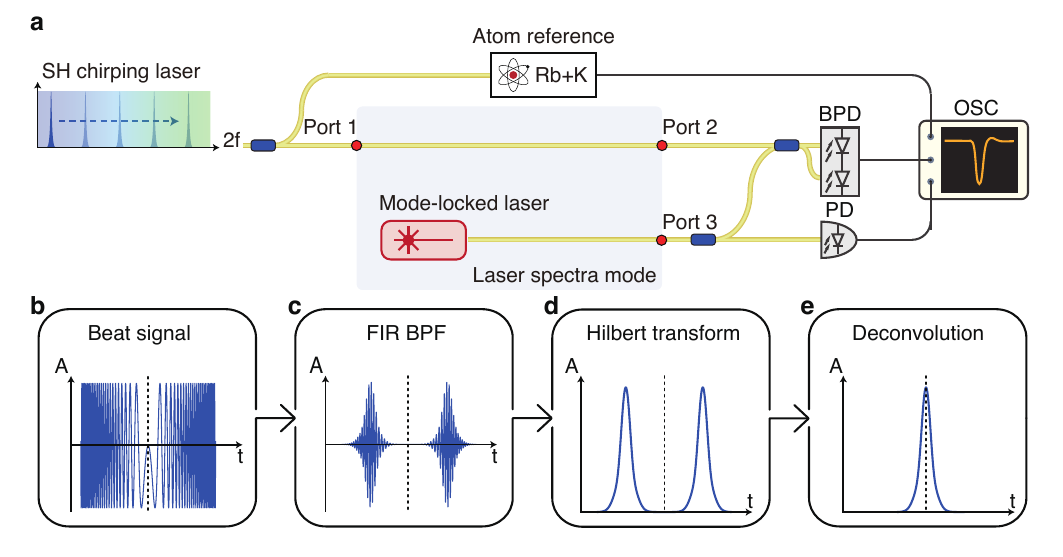}
\caption{
\textbf{Experimental setup and digital process to characterize mode-locked laser spectra}.
\textbf{a}.
Experimental setup.
\textbf{b-e}.
Flowcharts of digital process and algorithm. 
FIR, finite impulse response.
BPF, band-pass filter.
}
\label{SIfig:BPF}
\end{figure}
%%%%%%%%%%%%%%%%%%%%%%%%%%%%%%

%%%%%%%%%%%%%%%%%%%%%%%%%%%%%%%%%%%%%%%%%%%%%%%%%%%%%%%%%%%%
\vspace{1cm}
\section*{Supplementary References}
\bibliographystyle{apsrev4-1}
\bibliography{bibliography}